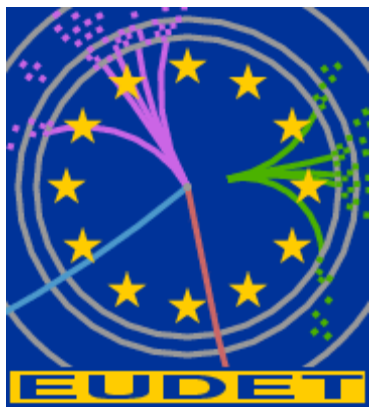

# Infrastructure for Detector Research and Development towards the International Linear Collider

## EUDET Consortium

## Abstract

The EUDET-project was launched to create an infrastructure for developing and testing new and advanced detector technologies to be used at a future linear collider. The aim was to make possible experimentation and analysis of data for institutes, which otherwise could not be realized due to lack of resources. The infrastructure comprised an analysis and software network, and instrumentation infrastructures for tracking detectors as well as for calorimetry.

J. Aguilar, P. Ambalathankandy, T. Fiutowski, M. Idzik, Sz. Kulis, D. Przyborowski, K. Swientek
*AGH Univeristy of Science and Technology, AGH-UST[a]*

A. Bamberger, M. Köhli, M. Lupberger, U. Renz, M. Schumacher, Andreas Zwerger
*Albert-Ludwigs Universität Freiburg, Germany[b]*

A.Calderone, D.G.Cussans, H.F.Heath, S.Mandry, R.F.Page, J.J.Velthuis
*Bristol University, UK[c]*

D. Attié, D. Calvet, P. Colas, X. Coppolani, Y. Degerli, E. Delagnes, M. Gelin[1], I. Giomataris, P.Lutz, F.Orsini, M.Rialot, F.Senée, W.Wang
*Commissariat a l'Energie Atomique, Paris, France[d]*

J. Alozy, J. Apostolakis, P. Aspell, F. Bergsma, M. Campbell, F. Formenti, H. Franca Santos, E. Garcia Garcia, M. de Gaspari, P.-A. Giudice, Ch. Grefe, V. Grichine, M. Hauschild, V. Ivantchenko, A. Kehrli, K. Kloukinas, L. Linssen, X. Llopart Cudie, A. Marchioro, L. Musa, A. Ribon, G. Trampitsch, V. Uzhinskiy
*European Organization for Nuclear Research, Geneva, Switzerland*

M. Anduze, E. Beyer, A. Bonnemaison, V. Boudry, J.-C. Brient, A. Cauchois, C. Clerc, R. Cornat, M. Frotin, F. Gastaldi, C. Jauffret, D. Jeans, A. Karar, A. Mathieu, P. Mora de Freitas, G. Musat, A. Rougé, M. Ruan, J.-C. Vanel, H. Videau
*Centre National de la Récherche Scientifique/Institut National de Physique Nucléaire et de Physique des Particules*
*Laboratoire Leprince-Ringuet Ecole Polytechnique, Paliseau, France*

A. Besson, G. Claus. R. de Masi, G. Doziere, W. Dulinski, M. Goffe, A. Himmi, Ch. Hu-Guo, F. Morel, I. Valin, M. Winter
*Centre National de la Récherche Scientifique/Institut National de Physique Nucléaire et de Physique des Particules*
*Institut de Recherches Subatomique, Strasbourg, France*

J. Bonis, S. Callier, P. Cornebise, F. Dulucq, M. Faucci Giannelli, J. Fleury, G. Guilhem, G. Martin-Chassard, Ch. de la Taille, R. Pöschl, L. Raux, N. Seguin-Moreau, F. Wicek
*Centre National de la Récherche Scientifique/Institut National de Physique Nucléaire et de Physique des Particules*
*Laboratoire de l'Accélérateur Linéare, Orsay, France*

M. Benyamna, J. Bonnard, C. Cârloganu, F. Fehr, P. Gay, S. Mannen, L. Royer
*Centre National de la Récherche Scientifique/Institut National de Physique Nucléaire et de Physique des Particules*
*Laboratoire de Physique Corpusculaire, Clermont Ferrand, France*

A. Charpy, W. Da Silva, J. David, M. Dhellot, D. Imbault, P. Ghislain, F. Kapusta, T. Hung Pham, A. Savoy-Navarro, R. Sefri.
*Centre National de la Récherche Scientifique/Institut National de Physique Nucléaire et de*

---

[1] Now at IreS

*Physique des Particules*
*Laboratoire de Physique Nucléaire et de Hautes Energies, Paris, France*

D. Dzahini, J. Giraud, D. Grondin, J. -Y. Hostachy, L. Morin
*Centre National de la Récherche Scientifique/Institut National de Physique Nucléaire et de Physique des Particules*
*Laboratoire de Physique Subatomique et de Cosmologie, Grenoble, France*

D. Bassignana, G. Pellegrini, M. Lozano, D. Quirion.
*Centro Nacional de Microelectrónica (CSIC)*
*Campus Universidad Autónoma de Barcelona, 08193 Bellaterra, Spain.*

M. Fernandez, R. Jaramillo, F.J. Munoz, I. Vila
*Consejo Superior de Investigaciones Cientificas, Madrid, Spain*

Z. Dolezal, Z. Drasal, P. Kodys, P. Kvasnicka
*Charles Univeristy, Prague, Czech Republic*

S.Aplin, O. Bachynska, T. Behnke, J. Behr, K. Dehmelt, J.Engels, K.Gadow, F.Gaede, E.Garutti, P.Göttlicher, I.-M. Gregor, T. Haas[2], H. Henschel, U. Koetz, W.Lange, V.Libov, W.Lohmann, B.Lutz, J.Mnich, C. Muhl, M.Ohlerich, N.Potylitsina-Kube, V, Prahl, M.Reinecke, P. Roloff[3], Ch. Rosemann, Igor Rubinski, P. Schade[4], S.Schuwalov, F.Sefkow, M.Terwort, R. Volkenborn
*Deutsches Elektronen Synchrotron, Hamburg and Zeuthen, Germany* [b]

J. Kalliopuska, P. Mehtaelae, R. Orava, N. van Remortel[5]
*Helsinki Institute of Physics, Helsinki, Finland*

J. Cvach, M. Janata, J. Kvasnicka, M. Marcisovsky, I. Polak, P. Sicho, J. Smolik, V. Vrba, J. Zalesak
*Institute of Physics, Academy of Sciences of the Czech Republic, Prague, Czech Republic*

T. Bergauer, M. Dragicevic, M. Friedl, S. Haensel, C. Irmler, W. Kiesenhofer, M. Krammer, M.Valentan
*Institute of High Energy Physics, Vienna, Austria*

L. Piemontese, A. Cotta-Ramusino
*Istituto Nazionale di Fisica Nucleare, Ferrara, Italy*

A. Bulgheroni, M. Jastrzab, M. Caccia
*Istituto Nazionale di Fisica Nucleare, Milano, Italy*

V. Re, L. Ratti, G. Traversi
*Istituto Nazionale di Fisica Nucleare, Pavia, Italy*

E. Spiriti

---

[2] Now at XFEL, DESY
[3] Now CERN
[4] Now at Zeiss
[5] Now at University of Antwerp

*Istituto Nazionale di Fisica Nucleare, Roma, Italy*

J.-P. Dewulf, X. Janssen[6], G. De Lentdecker, Y. Yang
*Inter University Insitute for High Energy Physics, Brussels*

L. Bryngemark, P. Christiansen, P. Gross, L. Jönsson, M. Ljunggren, B. Lundberg, U. Mjörnmark, A. Oskarsson, T. Richert, E. Stenlund, L. Österman
*Lunds Universitet, Sweden[e]*

S. Rummel, R. Richter, L. Andricek, J. Ninkovich, Ch. Koffmane, H.-G. Moser
*Max-Planck-Institute für Physik, Max-Planck-Gesellschaft, Munich, Germany[b] oben*

V. Boisvert, B. Green, M.G. Green, A. Misiejuk, T. Wu
*Royal Holloway and Bedford New College, Egham, UK*

Y. Bilevych, V.M. Blanco Carballo, M. Chefdeville, L. de Nooij, M. Fransen, F. Hartjes, H. van der Graaf, J. Timmermans
*NIKHEF, Amsterdam, Netherlands [f]*

H. Abramowicz, Y. Ben-Hamu, I. Jikhleb, S. Kananov, A. Levy, I. Levy, I. Sadeh, R. Schwartz, A. Stern
*Tel Aviv University, Israel*

M.J. Goodrick, L.B.A. Hommels, R. Shaw. D.R. Ward
*The Chancellor, Masters and Scholars of the University of Cambridge, UK*

W. Daniluk, E. Kielar, J. Kotula, A. Moszczynski, K. Oliwa, B. Pawlik, W. Wierba, L. Zawiejski
*The Henryk Niewodniczanski Institute of Nuclear Physics, Polish academy of sciences, Cracow, Poland AGH [a]*

D.S. Bailey, M. Kelly
*The University of Manchester, Manchester, UK*

G. Eigen
*University of Bergen, Norway*

Ch. Brezina, K. Desch, J.Furletova, J. Kaminski, M. Killenberg[7], F. Köckner[8], T. Krautscheid, H. Krüger, L. Reuen, P. Wienemann, R.Zimmermann, S. Zimmermann
*Universität Bonn, Germany [b]*

V. Bartsch, M. Postranecky, M. Warren, M. Wing
*University College London, UK[c]*

E. Corrin, D. Haas, M. Pohl
*Université de Genève, Switzerland*

---

[6] Now at Univ.Instelling Antwerpen
[7] Now at CERN
[8] Now at RWTH Aachen

R. Diener[9]
*Universität Hamburg, Germany*[b]

P. Fischer, I. Peric
*Universität Heidelberg, Germany* [b]

A. Kaukher[10], O. Schäfer, H. Schröder, R. Wurth
*Universität Rostock, Germany*[b]

A. F. Zarnecki
*Warsaw University, Warsaw, Poland*[a]

[a] Supported by the Polish Ministry of Science and Higher Education and its grants for Scientific Research
[b] Supported by Bundesministerium für Bildung und Forschung, Germany
[c] Supported by the Science and Technology Facilities Council, UK
[d] Supported by the Commissariat à l'Energie Atomique et aux Energies Alternatives, Saclay, France
[e] Supported by the Swedish Research Council
[f] Supported by FOM and NWO, Netherlands

---

[9] Now at DESY
[10] Now at XFEL, DESY

# Introduction

An electron positron linear collider has been recognized as the next big project with the highest priority by the high energy physics communities around the world. The choice of the superconducting technology for the International Linear Collider (ILC) had very much focused the communities and spurred activities at an unprecedented level on the accelerator side. To match this pace in the accelerator community, work on advanced detector concepts needed to be started, to be ready in time for a concrete proposal for a detector at the linear collider once physics results from the Large Hadron Collider (LHC) become available and indicate the energy range in which new particles are to be studied.

The project EUDET "Detector Research and Development towards the International Linear Collider" was an integrated infrastructure initiative funded by the European Union. It created a coordinated European effort towards research and development for the next generation of large-scale particle detectors and integrated within Europe the research activities, the infrastructures and the expertise of the participating institutions. New and advanced detector technologies are needed to fully exploit the potential of future accelerators like the ILC which is being designed in a worldwide collaboration. The project has developed infrastructures to facilitate experimentation and to enable the analysis of data using shared equipment and common tools. It thus established a common European infrastructure for the research on advanced detector concepts for the ILC and fostered collaboration between European partners and associated institutes.

While R&D on advanced detector technologies was already being pursued in several institutes their impact was limited by the lack of resources for coordination, networking and common infrastructure. Extensive R&D on detector concepts took place in the past in preparation for the Large Hadron Collider (LHC). The thrust and emphasis of that work was very different from the ones needed for the ILC, mainly with respect to granularity and radiation dependence.

The physics potential of a future linear collider represents a tremendous challenge for the performance of detectors, which is in most cases orthogonal to the main directions of development oriented towards LHC experiments. Rather than emphasizing radiation hardness and rate capability, the demands for precision and segmentation exceed significantly what is now state of the art. Vertex detectors should have 30 times smaller pixel sizes and be 30 times thinner, which is pushing the limits of sensor and electronics integration and lightweight precision mechanics at the same time. Large track detectors aim at 10-fold precision improvements while simultaneously requiring a reduction of the material budget by a factor of 6 at least and the particle flow approach to calorimetry drives the design of highly compact and granular devices with a channel density more than 2 orders of magnitude above the finest segmented LHC detectors.

Such leaps in performance cannot be achieved by simple extrapolation of the known, but only by entering new technological territory in detector R&D. Several new concepts for silicon sensor integration are being pursued for pixel devices, new micro-pattern gas amplification techniques are explored for use in time projection chambers, and calorimeter developers are pioneering the application of new solid state photo sensors for optical readout, to quote just some examples. While these new ideas have proven to be viable on desk-top scale or in typical test beam set-ups for proof-of principle studies, the demonstration of integration feasibility at large scales still has to be done. The R&D enters a new stage, which brings new infrastructure needs with it.

EUDET had been launched to address these in a coordinated manner. The research towards improving sub-detector specific infrastructure covered the following areas.

**Networking activities** addressed the need for a common software frame-work for all sub-detectors of a complex system, not only to avoid multiple developments but also to exchange and combine results in the R&D phase, to analyze data from combined beam set-ups, and to transfer the validated modeling of detector response into simulations. This is particularly true for the models of hadronic shower evolution, which acquired significant development momentum through EUDET and the close contact with test beam experimenters, and which is of relevance much beyond the linear collider community, namely also for the physics exploitation presently at full swing at the LHC.

**Test Beam Infrastructure** for the study of precision tracking devices. The test beam infrastructure consisted of a pixel telescope which defines the beam geometry to the precision adapted for the study of resolutions on the micro-meter scale, embedded in a mechanical structure which holds Devices Under Test (DUT) precisely in place.

**Infrastructure for Tracking Detectors**, which addressed large tracking devices. A superconducting magnet, Permanent Current Magnet (PCMAG), with a bore large enough to test a large TPC prototype (LC TPC), and a wall thin enough not to disrupt the incoming beam, was refurbished, and installed and precisely mapped. A large electrostatic field cage, adapted to the bore of the superconducting solenoid magnet, was built. It provides the extremely homogenous electrical field to transfer ionization charge to micro-pattern read-out structures, for which various innovative technologies are explored. Together with versatile mechanical interfaces and common read-out electronics, it has enabled many groups to advance their developments.

The main themes of the **Calorimeter** activity were compactness and electronics integration. The immense channel count and the fine segmentation of particle flow calorimeters must be realized in very dense tungsten and steel structures and requires high level integrated mixed circuit design respecting ultra-low power dissipation limits. Making use of common building blocks, a micro-chip family has been developed and used to equip compact mechanical structures, newly built within this activity, with versatile readout electronics, such that the novel sensor technologies can be tested and their functionality be validated under these extreme but realistic conditions.

## Networking Activities

An essential part of EUDET was the creation of a network of European institutions, which are participating in detector R&D for the ILC. The main objective of the networking activity was to strengthen the European part of the worldwide efforts towards a detector at the ILC by providing a specific set of tools and structures. These tools and structures were provided to all participants of EUDET and furthermore to all interested parties in Europe and around the globe. There were three major aspects of this network. First, a human network was created through the organization of five Annual Scientific Workshops and through enabling travel between the partners and associated parties. Second, common analysis of both test beam and simulation data sets was enabled through the provision of additional computing hardware and a common software framework, iLCSoft [5], which was substantially supported by EUDET.

In addition, substantial improvement was made in a specific critical item of ILC detector research, the modeling of showers of hadrons in fine-grained calorimeters within the world-leading simulation suite Geant4. Third, facilitated access to commercial deep sub-micron electronics technologies for radiation tolerant microelectronics developments for front-end and readout ASICs was provided including dedicated training courses for microelectronics developers at the partner institutes.

Indeed, all tasks within the networking activities provided crucial instruments for the accomplished R&D successes in the three detector developments areas listed above, which are reported in this document, as well as for the successful completion of the Letters of Intent for the ILC detectors. Also, the current process of producing Detailed Baseline Documents for the ILD and SiD detectors as well as the CLIC Conceptual Design Report, both due in 2012, is profiting significantly from these structures.

These structures and their essential elements had to be available early in the course of the project, in order to be used within the EUDET activities. After that, the focus moved towards further improving the tools and supporting the use of the networking instruments. In the following, the objectives and accomplishments of the six tasks within the networking activity are summarized.

## 1.1 Computing Infrastructure

In the course of the EUDET project, a significant amount of test beam measurements have been performed in all three EUDET detector activity areas as well as by external users of the infrastructures. Several tens of Terabytes of test beam and simulation data were produced. In order to be able to access these data remotely from all over the world and to provide distributed computing resources for their analysis, EUDET adopted the GRID paradigm early on. While in the beginning of the project, the worldwide computing grid for high energy physics (HEP) was still in its infancy, in the course of the project grid computing has become the generally adopted computing model in HEP. Dedicated storage and computing resources for EUDET have been set up at Bonn University, DESY and Tel Aviv University. The resources were commissioned in the first two years of the project and are now routinely used since 2008. They are integrated fully transparently in the ILC and CALICE Virtual Organisations (VOs) to which they contribute approximately 10%. Several hundred million events in several ten thousand runs of the EUDET beam telescope are stored and have been analyzed. All data of the large TPC prototype are available and their analysis is still ongoing. Several hundred million calorimeter beam test data have been stored. Approximately 100.000 normalized CPU hours per month are provided through this task.

## 1.2 Analysis Software Framework

The reconstruction and analysis of data from test beam campaigns are complicated and usually time and person power consuming. It is in the nature of test beam data that their format and contents is rapidly changing with time and different for different detectors, user groups etc. Without careful planning, large overhead is easily created by performing similar tasks like formatting data, reconstructing objects, matching objects from different sub-detectors etc. over and over again for each new setup. In order to significantly reduce this overhead and to provide the test beam data in a durable, standardized and well-documented format to the different user groups, a common data analysis and simulation infrastructure is vital. With the help of this task, the iLCSoft software framework has been developed and made usable for the analysis of test beam data [1].

The objective of the task is the development of a common data analysis and simulation infrastructure for the ILC detector R&D. The iLCSoft software framework was originally developed for studying the Monte-Carlo simulation of a full detector design for the ILC. It has been subsequently extended and adapted for test beam data processing within the EUDET project. The framework provides the core software tools for the development of reconstruction and analysis software, and allows these to be used in conjunction with Mokka, a Geant4 based detector simulation application, through the provision of the LCIO Event Data Model (EDM) Application Programming Interface (API) and Geometry API for Reconstruction (GEAR) [2].

The basis of all software tools is the LCIO event data model and persistency framework [3]. It is already fairly complete and flexible. Nevertheless its data structures have been extended by classes, which are specifically suited for raw data as needed in the prototype studies on request by the user community. It is complemented by GEAR to describe the detector geometry in a standardized and flexible way. The backbone for analyzing and reconstructing data is the Marlin application framework [4]. Its modules, called processors, sequentially process the event data by reading data from an input collection, process it and fill the results in an output collection, which in turn can be processed by a subsequent processor. User control is provided through XML files. The modularity and well-defined interfaces between the processors ensure that different developers can work on different processors in parallel without interference.

The Linear Collider Conditions (LCCD) Toolkit is used for storage and retrieval of conditions data via time stamps. iLCSoft is completed by many tools that increase the usability and efficiency for the users, such as the build tool ilcinstall, the Marlin Graphical User Interface (GUI) or the C Event Display (CED). All software tools are fully functional since the first release version 1.0 of the iLCsoft in 2007. The three detector activities of EUDET have fully adopted the common software framework and have based their specific analysis packages (CaliceSoft, MarlinTPC and EUTelescope) on it. The CLIC study group has adopted iLCSoft for writing their CDR. A web portal has been set up to provide a common entry point to documentation and download links for all software packages as well as to the web-based access to the central source code management system [5]. It also offers the users links for downloading the software and for online browsing of the subversion (svn) code repositories. The software is fully compatible with using it within the grid.

The iLCSoft framework continues to be maintained after the end of EUDET and provides a highly developed and generally adopted tool for the study of advanced detectors for linear colliders e.g. within the FP7 initiative AIDA.

## 1.3   Simulation of Hadronic Showers in Geant4

The goal of this task was to improve the modeling of hadronic showers within the Geant4 toolkit. This improved modeling is necessary in order to fully exploit the fine-grained particle flow calorimeters, which are studied for the linear collider detector concepts. Also, tools for the validation of these improvements based on test beam measurements from the CALICE collaboration and other experiments have been provided [6].

Over the course of the project, various improvements of the different physics models (so-called physics lists) that are available within Geant4 have been performed. The most recent version of Geant4, version 9.4 includes all recent developments. A comparison of real data from CALICE test beam measurements with the improved simulations was performed and especially extensive comparisons of hadron data from the CALICE collaboration with the most recent Geant4 physics lists were presented. New observables were reported, which allow for an unprecedented level of detail in the comparisons. These observables include energy profiles after the identified interaction point and the response in sections of the calorimeter in the first few layers after the interaction, in the next vicinity (where photons would interact) and in distant layers. Overall, reasonable agreement was seen; however deficiencies in energy deposition in the layers near an interaction and the subsequent rise were clearly observed. These have provided feedback for further improvement of the models.

## 1.4   Microelectronics Support

The development of Application Specific Integrated Circuits (ASICs) is vital to meet the readout challenges of detectors for linear colliders and other future HEP experiments. Integrated circuits with feature sizes in the deep submicron regime allow for the implementation of fast and efficient data reduction and extremely low-noise and low-power electronics. Deep submicron electronics has also been shown to be inherently radiation tolerant. In order to facilitate access to industrial deep submicron technologies, the MICELEC task was created within EUDET. The objectives within this task were to give design support and coordinated access of the ILC community to a commercial silicon foundry for prototyping and production of integrated circuits.

Based on a contract for a 250 nm technology that was previously available at CERN for the construction of electronics for the LHC experiments and which will continue to be supported, a new foundry contract on 130 nm CMOS technology has been negotiated with a vendor preceded by substantial work of technology validation. Various ASICs, which have been developed in the context of EUDET, were supported through this task. Among these were: a low-noise programmable charge amplifier (PCA16) for a new read-out system for a Time-Projection-Chamber; a read-out chip for a pixel-based TPC (TimePix); various chips for a silicon tracker for ILC detectors; a low-power 10 bit, 40 MHz Analog-to-Digital (ADC) converter; an energy-optimized new filtering (digital signal processing) block and also of a low-power data compression block. The latter blocks entered into the 16-channel S-Altro TPC readout chip. Support is ongoing for the preparation of the testing, characterization and for the packaging of this complex chip.

Modern technologies, while offering unprecedented potentials to the design engineer, also are requiring the utilization of much more advanced design tools to support the complexity offered by the much higher level of integration available. A set of specific integrated tools has been acquired using both EUDET and CERN funds. This allowed creating a set of software tools to facilitate the digital design flow, which was distributed to the EUDET community. Several advanced training courses for microelectronics design engineers took place in the course of the EUDET project with high-priority access for EUDET partners.

## Test Beam Infrastructure

Within the EUDET project the improvement of test beam infrastructures has evolved. The test beam infrastructure consists of a pixel telescope, which defines the beam geometry very precisely. The design goals included high position resolution ($\sigma < 3.0$ μm) even at low momentum, and readout rates of 1 kHz. Highly granular and thin detection planes were required. At the time of approval no pixel telescope was available within the worldwide community. The construction was planned in two stages, to quickly offer an exploitable infrastructure in parallel to the development of the final telescope. In the first stage (from now on referred to as Demonstrator) a well-established CMOS pixel technology with an analogue readout was used. The analogue-to-digital conversion and signal processing was realized using fast processors in the readout electronics. This Demonstrator did not satisfy the final requirements with respect to chip size and readout speed, but a first test facility was available quickly to satisfy immediate and urgent test needs of various research groups working on pixel detectors in Europe. From summer 2007 until fall 2010 numerous High Energy Physics detector R&D groups used the Demonstrator telescope for their test beam studies at the DESY electron beam and the CERN-SPS hadron beam. In the second stage of the project the Final Telescope was designed and commissioned. In this report the design of the Final Telescope, the construction, and the performance will be summarised.

### 1.5 Pixel Telescope: Motivation and Overview

During the R&D phase for particle detectors a number of beam tests are needed to show the performance of the newly developed devices. In order to extract parameters such as resolution and efficiency, the track of the beam particle needs to be defined precisely, and usually beam telescopes are used for this purpose. Different telescopes have been built in the recent decades; examples include the Burst Alert Telescope (BAT) in the Swift Satellite and the Micro-Vertex Detector (MVD) telescope at ZEUS. As these telescopes are built for a specific DUT, their use by a different R&D group usually results in a lot of adjustment work such as rewriting the data acquisition code. The idea of the EUDET pixel telescope was to provide an easy-to-use system with well-defined interfaces enabling test beam studies on a rather short time scale. The telescope had to provide the precision for high momentum beams of electrons, pions, and protons at hadron machines as well as at low momentum electron beams at DESY (1-6 GeV), where the precision is dominated by multiple scattering.

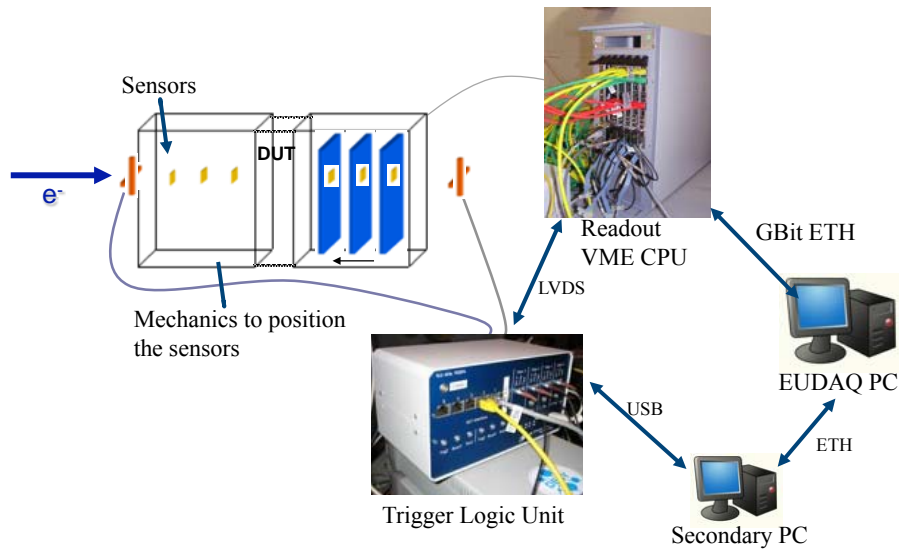

Figure 1: Overview of the components necessary for a beam telescope.

In Figure 1 all main components for such a pixel telescope are shown:
- Precise mechanics to position the sensor boards in the particle beam;
- Pixel sensors with a single point resolution of 2 - 3 μm and a minimum of material;
- Custom made readout boards adapted to the pixel sensors;
- A DAQ system including
  - Trigger Logic Unit (TLU)
  - DAQ computer and software framework;
  - Analysis software for reconstruction, alignment and tracking.

In the following sections the design of the telescope will be described including some technical details for each part of the system.

**Design Issues**

The beam telescope is to be used for a wide range of R&D applications and quite different devices under test (DUT), from small (a few millimetres) to large (up to one meter) size. Depending on the project and on the size of the device the requirements with respect to precision as well as coverage are quite different. The mechanical setup must therefore allow for a wide range of different configurations from a very compact layout useful for pixel sensors to a two-arm layout with sufficient adjustable space in between the arms to accommodate a large DUT. The lateral dimensions of the active area should be large enough to cover high precision pixel devices without mechanical movement of the device under test. Mechanical actuators adjust the DUT position within the telescope or scan the DUT in case of larger devices. Finally, the overall setup of the telescope must be flexible enough to make it transportable to other beam lines outside of DESY, e.g. to higher energy hadron beam lines at CERN. Still with a careful optimization of the telescope setup with respect to dead materials and positioning of the telescope planes, the precision of the predicted impact position of beam particles on the DUT plane can reach less than 3 μm even at 6 GeV.

A dedicated study was performed to understand the position resolution in the telescope, in order to optimize its performance by choice of the best plane setup. The approach is based on a novel analytic track fitting method, taking into account multiple Coulomb scattering effects. A schematic view of possible plane configurations of the telescope is presented in Figure 2. In these configurations, the telescope consists of three standard reference planes (in red) with 2 μm intrinsic resolutions, and one high-resolution reference plane (in blue) in the middle with 1 μm intrinsic resolution. In order to minimize the multiple scattering effect on the track measurement error, the high-resolution plane should be placed as close as practically possible to the DUT (in pink). The thickness of each plane should also be minimized as much as possible. CMOS sensors allow reducing the latter to a few tens of micrometers.

Based on these studies it was decided to provide six telescope planes, also for redundancy and flexibility. The telescope was subdivided into two arms of three sensors each, allowing also larger DUTs to be located between the two arms. A high-resolution plane was also provided to improve the resolution to ~1 μm.

The speed of the device allowed taking full advantage of the beam rates operating at readout rates of up to 1000 frames/sec.

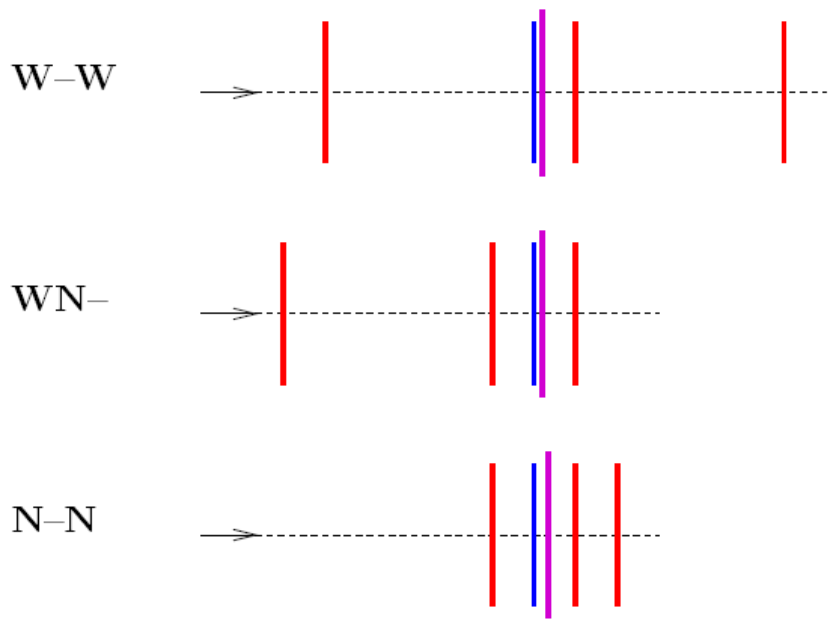

Figure 2: Possible plane configurations for the telescope

**Mechanical Setup**

Two telescope arms housing the reference planes are the main elements, leaving space between the arms for the DUT. Both arms are adjustable in the beam direction to ease the installation of the DUT as well as to give the possibility of larger DUT devices. DUTs up to 50 cm in size can be installed. For the DUT a high precision XYϕ-table can be used to move the device through the active area of the telescope. The table consists of a vertical, a horizontal, and a rotation stage.

Each sensor is positioned inside a sensor jig providing precisely machined pins for the exact positioning of the sensor. Three sensor jigs are guided on a rail system on each reference arm. The rail system allows a distance of up to 150 mm between the outermost planes. The minimal distance between the sensors is 20 mm. On the top sides of the jigs the auxiliary boards (AUX) are mounted. Special openings are provided for the cables to connect the sensor boards to the AUX boards. Holes where the beam passes through are closed with 25 μm thin Kapton foil. The boxes and the XYϕ-table are positioned on a support system built out of construction profiles[11]. This allows an easy modification when needed for the mechanical integration of the different DUTs. The lowermost part of the table is divided in two levels where the upper of the two is rotatable by a few degrees to ease the adjustment of the system to the beam axis.

A photograph of the setup installed at the DESY test beam is shown in Figure 3. The overall mechanics is not as compact as envisaged at the time of proposal, but in order to accommodate all the different mechanical needs from the users, a lot of flexibility had to be implemented. At the same time the mechanics was built to be very sturdy and stable over time.

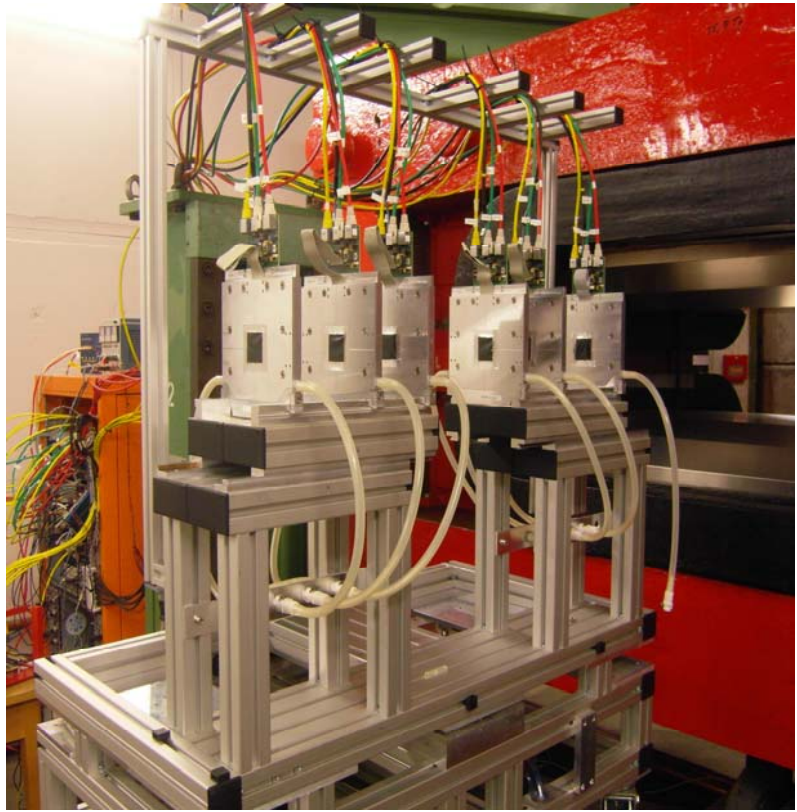

Figure 3: The EUDET Telescope installed at the DESY test beam.

**Pixel Sensors Equipping the Telescope**

The sensitive elements equipping each plane of the telescope have to be highly granular and thin. The performance achieved with CMOS pixel sensors at the beginning of the decade [7] indicated that this technology was already mature enough to fulfill the EUDET telescope requirements in terms of spatial resolution, material budget, sensitive area dimensions and readout time.

CMOS sensors are manufactured in standard processes by commercial CMOS microelectronics foundries. They allow detecting minimum ionising particles (mips) by collecting the charge liberated when traversing the thin, almost undepleted, epitaxial layer implemented on the wafer substrate. The carriers of the signal charge (electrons) diffuse thermally in this layer and are collected by sensing elements formed by regularly implanted n-wells in direct contact with the (p-type) epitaxial layer [8]. Since mips generate typically ~80 electron-hole pairs per micrometer in the 10 – 15 μm thick epitaxial layer, the signal charge ranges from a few hundred to ~1000 electrons.

---

[11]   Rose & Krieger

The sensors offer remarkable performance in terms of intrinsic resolution, material budget and readout electronics. The excellent intrinsic resolution follows from the high granularity achievable, combined with the cluster charge sharing due to thermal diffusion. The low material budget comes from the thin sensitive volume, which enables thinning the sensors down to a few tens of micrometres. Finally, the possibility to integrate read-out micro-circuitry on the same substrate as the sensitive volume results in a very low read-out noise, and in a powerful and flexible signal processing capability.

The sensors implemented in the telescope were manufactured in a CMOS process with 0.35 μm feature size and a 14 μm thick epitaxial layer. Their design was guided by simplicity and robustness. The micro-circuitry integrated in each pixel is therefore restricted to two transistors and a forward biased diode. The transistors are used to connect the pixel to a source follower integrating the charge collected, and for addressing the pixel at read-out and signal transfer. The self-biased diode ensures the reverse biasing of the sensing diode and provides a continuous compensation of the leakage current it delivers.

The telescope reference planes are equipped with 18.4 μm pitch sensors, called Mimosa26. Mimosa26 is the first pixel sensor covering an active area of ~224 $mm^2$ with integrated zero suppression. A single point resolution better than 4 μm is achieved with a pixel pitch of 18.4 μm. The matrix is organised in 576 rows and 1152 columns. At the bottom of the pixel array, each column is connected to an offset compensated discriminator to perform the analogue to digital conversion. Digital data are then treated by a zero suppression circuit in order to send useful information. This architecture allows a fast readout frequency of ~10 kframes/s. The reference planes are complemented with high-resolution sensors, called MIMOSA-18, featuring a 10 μm pitch.

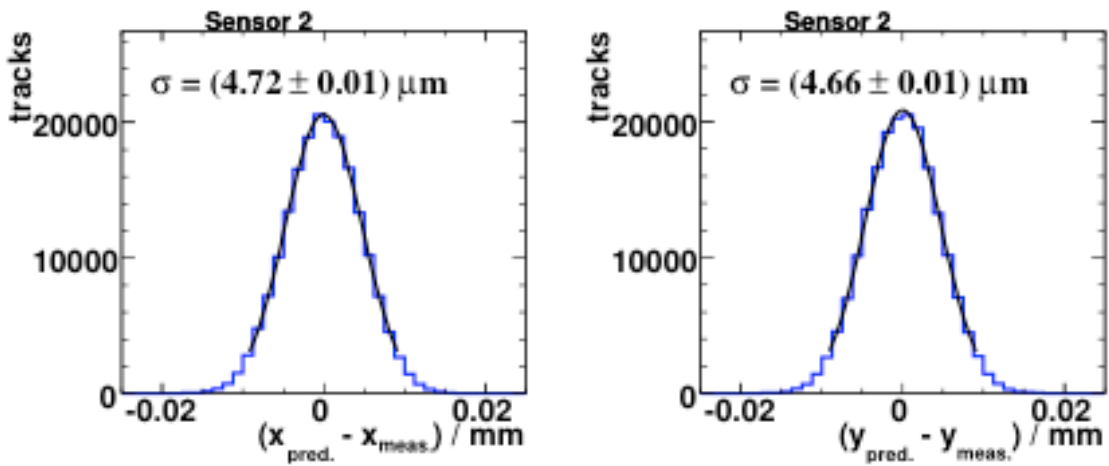

Figure 4: Residual distribution in x and y for MIMOSA-26 sensors with S/N of 10

The residual distributions for one of the MIMOSA-26 sensors are shown in Figure 4. The intrinsic telescope resolution calculated from these results (using only 4 telescope planes) was found to be 2.22 μm. However, it is significantly improved when using all 6 planes, and by lowering the S/N threshold, to give a resolution better than 2 μm.

## 1.6 The DAQ System

The data acquisition system, as shown in Figure 5 is a combination of dedicated hardware and software with the following sketched data flow: the pixel sensors are read out by custom made data reduction boards (EUDET Data Reduction Board - EUDRB [9]), and data are sent via VME64x block transfer to a VME CPU. The data are then packaged and sent over Gigabit Ethernet to the main DAQ PC where the events are built and written to file.

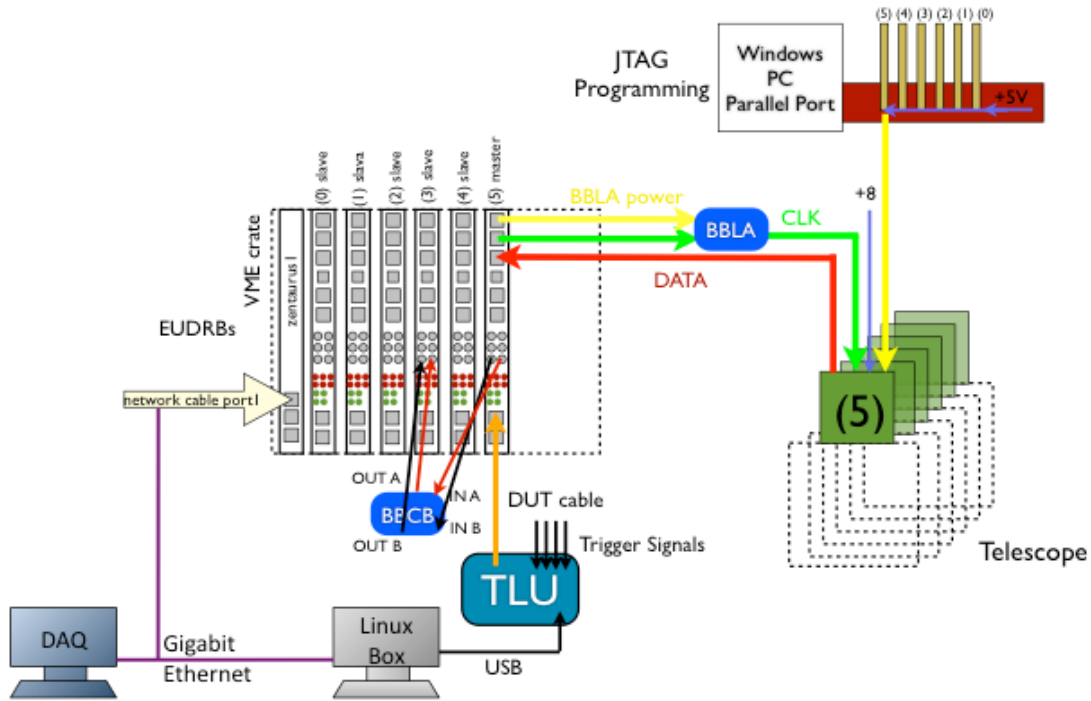

Figure 5: Overview of the telescope DAQ system

**The Trigger Logic Unit**

Triggering is controlled by a custom Trigger Logic Unit (TLU) (Figure 6), that receives signals from scintillators in front of and behind the telescope, and generates triggers that it distributes to the telescope and any DUT. The four scintillators inputs can be combined in an arbitrary fashion (AND, OR, VETO) to generate the trigger signal, and there is also an internal trigger generator for testing and pedestal runs. Several internal scalers allow monitoring of the running conditions. For each generated trigger a trigger counter is incremented, and a timestamp is stored in an internal buffer that may be read out over USB by a PC. The DUTs have the option to read out the trigger number via the DUT interface in order to ensure proper synchronization of triggers.

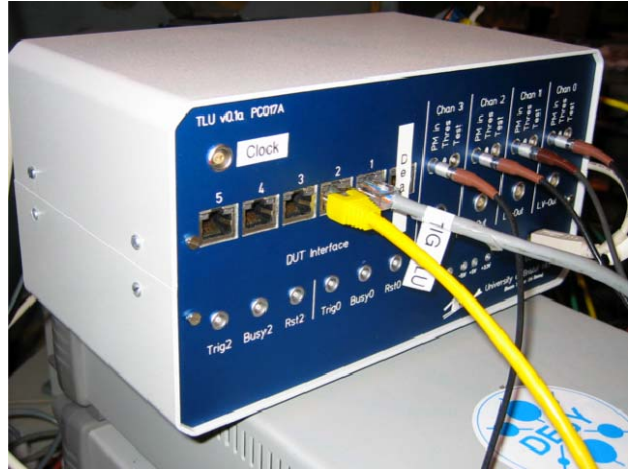

Figure 6: The TLU

Up to 6 devices (including the DUT and the telescope) may be controlled by the TLU. DUTs can connect either via the RJ45 connectors using LVDS, or via the LEMO connector using either NIM or TTL levels. In addition, they have a choice of handshake modes. In the simplest mode, just a trigger pulse can be sent to the DUT without any handshake. For slightly improved robustness, the DUT may respond to each trigger with a busy signal, dropping the busy when it is ready for another trigger. This ensures no triggers are dropped since the TLU is expecting a response from the DUT. The recommended mode is similar to the trigger-busy mode, with the difference that during the busy period the DUT toggles a clock line, and the TLU responds by sending the trigger number bit by bit over the trigger line. This allows any desynchronisation of triggers to be recovered offline in case anything goes wrong.

**The EUDRB Data Reduction Boards**

The sensors are read out by custom readout electronics. Originally designed for the analogue sensors of the Demonstrator telescope, they have been adapted to read out the fully digital MIMOSA-26 sensors. The EUDRBs, shown in Figure 7, are built on 6U VME64X cards, with the main processing being carried out by an Altera Cyclone II FPGA that includes a NIOS II soft microcontroller used for diagnostics and configuration.

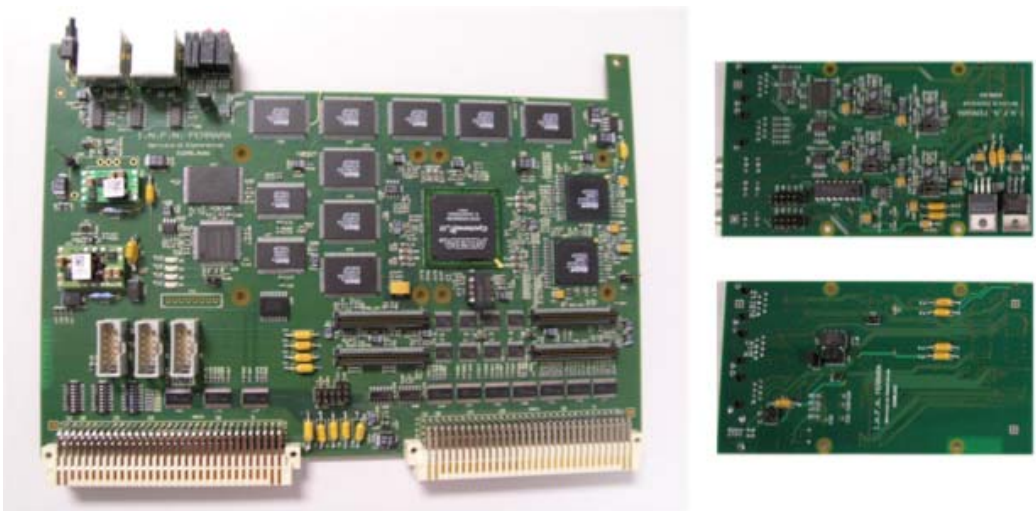

Figure 7: The EUDRB (left), and the analogue (upper right) and digital (lower right) daughter cards

The interface to external electronics is provided by two daughter cards, one for analogue and the other for digital signals. This modular design has allowed the EUDRBs to be upgraded to handle the new MIMOSA-26 sensors without a complete redesign.

**Data Acquisition Software**

The Data Acquisition software, EUDAQ, is a custom framework designed to be lightweight, modular and portable. The overall architecture is shown in Figure 8. A central Run Control displays a graphic interface to the user allowing them to control the other processes. The Data Collector receives data from the Producers, merging the separate streams into one and writing it to file. The Producers communicate with the different pieces of hardware and generate data. For example, the TLU Producer controls the TLU, and sends the timestamps to the Data Collector, while the EUDRB Producer controls the readout boards and sends the telescope data.

In addition, a Log Collector collects log messages from all other processes into a central location so that they may be read more easily. The online monitor reads the data file from disk, generating histograms so that the user may monitor the quality of the data.

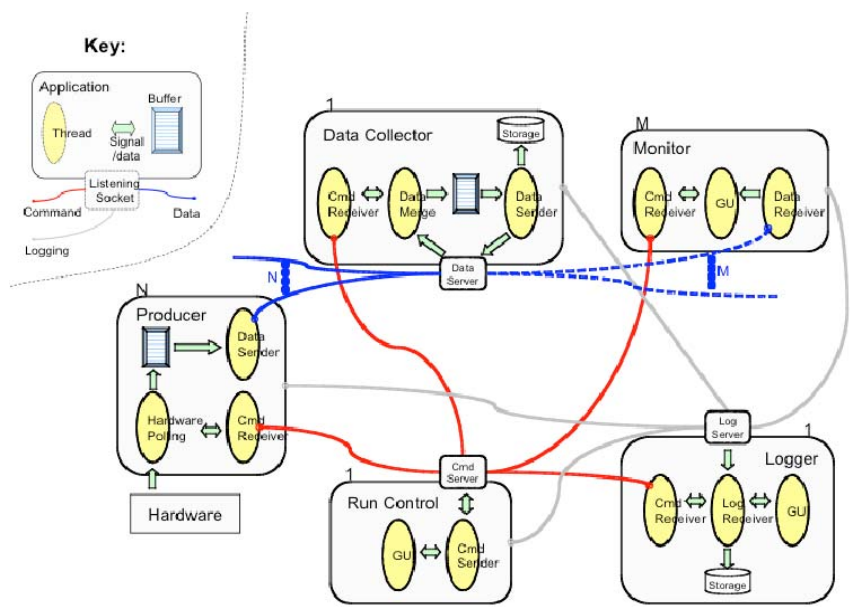

Figure 8: Overview of the DAQ software architecture

## 1.7 User Integration

The telescope is meant to be used by many different user groups with widely varying needs; therefore it is designed to be as flexible and user friendly as possible. The TLU can communicate with DUTs using one of three different voltage levels, and a choice of protocols, from just a simple trigger pulse with no handshake to sending the trigger number with each trigger, allowing the user to choose the best compromise between simplicity and robustness.

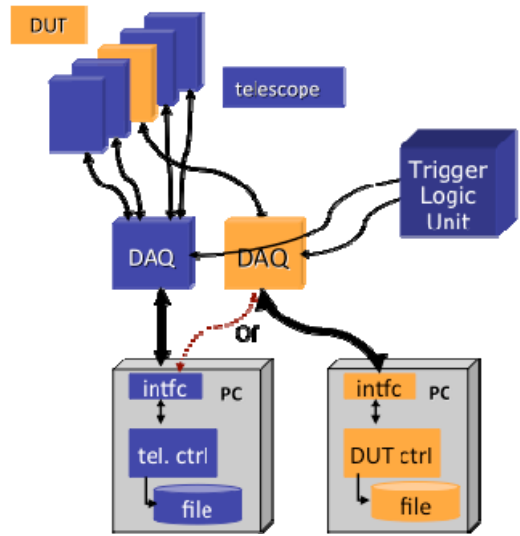

Figure 9: DUT integration with the telescope

A schematic showing the different levels of integration on the software side is shown in Figure 9. At the simplest level, the DUT only communicates with the TLU, and has a completely separate DAQ. This method requires the least amount of preparation, but since the two DAQs are completely separate, starting and stopping runs must be synchronised manually, and run numbers must also be matched up manually.

In order to alleviate these problems, an intermediate level of integration was developed, where the user still keeps a separate DAQ system, but it is controlled by a Run Listener that receives commands from the telescope Run Control, meaning there is only one place to start and stop runs, and the run numbers can be automatically synchronized.

However, having two separate DAQ systems means that the online monitoring systems are also separate, so it is not possible to correlate data between the DUT and the telescope. With full integration, the user writes a Producer for their DAQ, so that their data can also be sent to the central DAQ PC and merged with the telescope data. The EUDAQ software has been designed with this in mind, so it is relatively easy to integrate new DUTs into the system, usually taking only a few days work with the help of an EUDAQ expert. With full integration, the online monitor is able to access all of the data and generate correlation plots between the DUT and the telescope, giving fast feedback of the position of the DUT with respect to the telescope.

## 1.8 Performance

The single layer resolution during beam tests was determined to be between 3.5 μm and 4.5 μm, depending on the threshold settings, in both horizontal and vertical direction. This leads to a typical resolution of the interpolated space point on a DUT of less than 2 μm, when the DUT is placed in the center of the two telescope arms. Massive DUT can be accommodated behind the telescope, with extrapolated space point resolutions of less than 5 μm when the DUT is placed as close as possible. Even at distances of 1.5m behind the telescope, the space point prediction is accurate to 25 μm. The operation of the telescope was stable and required little human intervention. Trigger rates of the order of 1 kHz were regularly achieved, in fulfillment of the original specifications for the device.

Starting in 2011, the telescope will be the basis of further developments for the FP7 program AIDA. Within this program, one of the infrastructures is a beam telescope for the characterization of pixel detector prototypes, thus a logical continuation of the EUDET telescope and its surrounding infrastructure. The new telescope will have to cater to sLHC needs (requiring a $CO_2$ dual phase cooling plant and fast telescope arms). Infrastructure for thermo-mechanical characterization is also envisaged at DESY. Starting on the basis of the present telescope, a next generation device is being planned. User will be able to choose from three different technologies: ATLAS pixel detectors with LHC timing; Timepix high precision timing and high resolution; and large area, high resolution Mimosa MAPS. A segmented trigger will provide easier tracking using a scintillator hodoscope. An advanced version of the TLU will provide track tagging in case of high rates.

# Infrastructure for Tracking Detectors

A large Time Projection Chamber (TPC) is one of the proposed options for the central tracking system of a detector at a future $e^+e^-$ linear collider [10]. It provides continuous tracking over a large volume, with many (of order 200) space points along a particle track. The TPC needs to have a thin field cage, provide excellent single hit resolution and double track separation and this can be achieved using micro-pattern gas detectors (MPGDs) like GEMs or Micromegas as gas amplification systems. The envisioned single point resolution in the RΦ projection is 100 μm or better over the full drift space in a magnetic field of 3.5 T, leading to a transverse momentum resolution of $\delta(1/p_t) \leq 9 \times 10^{-5}$ $(GeV/c)^{-1}$ for the TPC alone.

In order to test the technologies necessary to achieve these goals a Large Prototype TPC was constructed in a combined effort with the LCTPC collaboration [11] and mounted in a 1 Tesla solenoid magnet PCMAG (see section 4.1), at one of the DESY electron test beams.

## 1.9 Superconducting Magnet PCMAG

The superconducting magnet PCMAG (Persistent Current MAGnet) was provided by KEK. PCMAG is a rather lightweight magnet: a low mass coil and no return yoke make its weight to be 460 kg. It has a usable diameter of about 85 cm and the usable length is about 130 cm. Its 3342 windings with an operating current of 480 A provide a magnetic field density of up 1.25 T in the centre region of the magnet. The field is homogeneous within 3% in the region of 30 cm around the centre, whereas larger deviations are expected in the remaining region (Figure 10). This, however, allows for establishing correction tests during operation.

The device was delivered to DESY in 2006 and first time running at the DESY test beam was achieved in the same year. In the following year the infrastructure for the helium transfer line was improved for easier use. The magnet was operable throughout the rest of the project and regularly used by user groups. A field measurement was completed and the proper determination of the field map was performed [12].

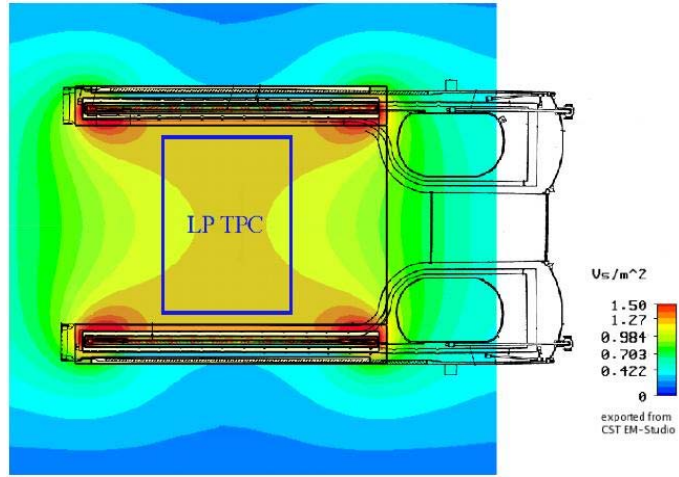

Figure 10: Schematic drawing of PCMAG and its field density distribution

## 1.10 The Large Prototype TPC

A rather complete setup has been established at the DESY test beam, providing an infrastructure for a world-wide effort in the development of a large TPC to be used as main tracking device at a future linear collider. It consists of the following items:

1) Large scale ($\varnothing \approx 1$ m), low mass field cage
2) Modular end plate system for large surface GEM and Micromegas systems
3) MPGD detector modules
4) Prototype readout electronics
5) Infrastructure: test beam, magnet, supporting devices, HV and gas systems, slow controls
6) Silicon envelope detectors
7) Software developed within the MarlinTPC [13] framework for simulation, calibration and reconstruction of TPC data

The LP has a diameter of 770 mm and a length of 610 mm and fits into the superconducting magnet PCMAG. Tracks measured within the LP can have up to 125 space points, using anode pad readout, depending on the pad size. The aim of these tests is not only to confirm the anticipated single point resolution, but also to address issues related to the large size of this TPC, like alignment, and electric and magnetic field distortions.

**The Field Cage**

A composite field cage structure was developed which combines light-weight and excellent high voltage behaviour and mechanical stiffness [14]. In addition a holding structure for the field cage was developed which allows the manipulation of the field cage inside the magnet without putting excessive strains on the structure. The homogeneous electric drift field for electrons/ions was provided by a series of field strips and their 'mirror' strips at an intermediate potential, such that an electric field uniformity at the level of $10^{-4}$ should be reached. The field cage can be seen in Figure 11. After survey it appeared that the mechanical axis of the barrel deviated by 500 μm from the electric field axis, a factor 5 more than the tolerance, leading to a non-uniformity of the electric field $\Delta E/E$ between $10^{-4}$ and $10^{-3}$.

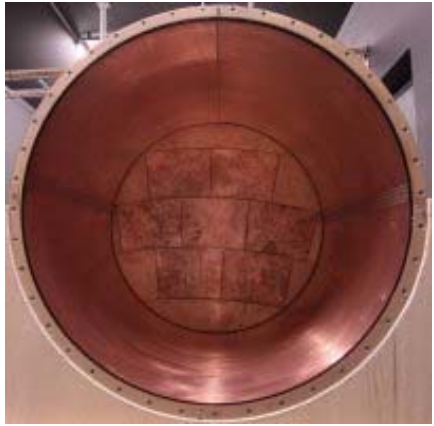

Figure 11: The LP field cage. On the back side one can see the inner part of the anode end plate equipped with seven dummy modules.

**Anode Endplate**

The gas amplification modules can be mounted in the anode end plate [15] in a pattern that is a circular subsection of a possible TPC at a future linear collider. The end plate allows positioning the modules with an accuracy of better than 50 μm. Several areas have been cut out in order to implement further 'service' devices for usage with the TPC, e.g. laser insertion holes. Figure 12 shows both a schematic picture of the inside view of the end plate and an outside view with modules mounted.

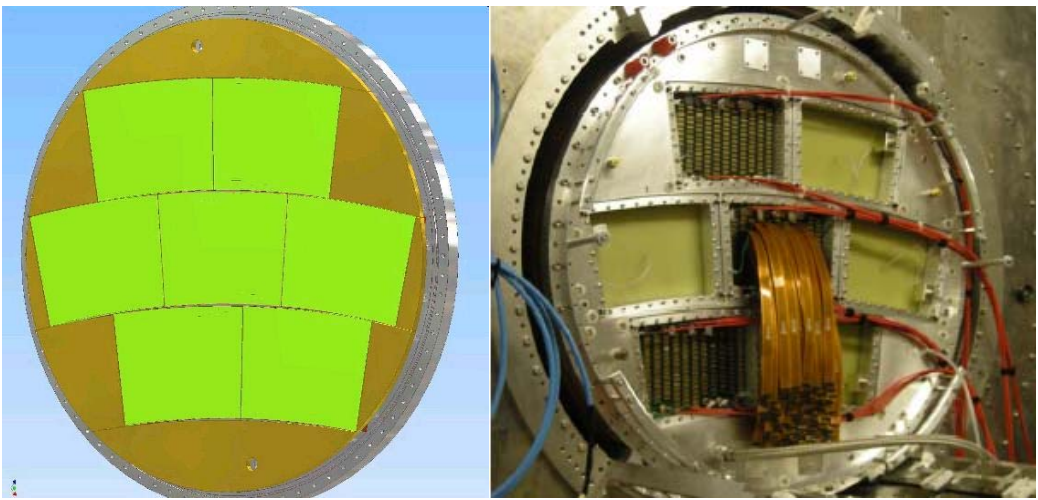

Figure 12: Schematics of the anode end plate with amplification modules and termination plates (left). View of the LP TPC mounted inside PCMAG. Three active double-GEM endplate modules are mounted (right).

## 1.11 Amplification Modules

The LP TPC has been equipped with several types of MPGD gas multiplication structures with small pads (few $mm^2$) as charge collecting anodes. Single Micromegas [16] modules, without and with resistive layer [17], covering the pads of size $3 \times 6.8$ $mm^2$ was used, as well as up to three modules with two or three GEM [18] foils with naked pads. The pad size in case of the two layer GEM system was $1.1 \times 5.25$ $mm^2$. The resistive layers in case of the Micromegas are needed because of otherwise insufficient spreading of the collected charge over only 1-2 pads, thereby deteriorating the single point resolution. Also CMOS pixel readout ASICs (Timepix) have been successfully used on endplate modules, both with GEMs and with Micromegas-like (Ingrid) gas multiplication (see section 4.7). In the near future modules will be equipped with gating devices in order to reduce the ion backflow into the drift volume.

## 1.12 Readout Electronics

In case of the GEM modules the pad readout is based on the readout electronics that was developed for the ALICE experiment at the CERN LHC, the ALTRO system [19-20]. The 16 channel ALTRO chip performs analogue to digital conversion with 10 bit precision followed by various steps of digital signal processing, including zero suppression and storage in an event buffer. The sampling can be clocked at frequencies up to 40 MHz so in principle sampling at this frequency and frequencies lower by multiples of two is possible. However, at 40 MHz sampling the full resolution is not maintained for the standard ALTRO chip. A limited number of ALTRO chips were modified to allow sampling at 40 MHZ with almost full precision. Up to now the system has been operated at 20 MHz only. The ALTRO chip has an event storage memory of 1 k 10-bit words, which corresponds to sampled data over a depth of 50 μs drift time at 20 MHz sampling frequency. The so called T2K gas mixture (95/3/2 % Ar/$CF_4$/Isobutan) was used in the TPC, which gives a drift velocity of around 7 cm/μs at a drift field of 230 V/cm. This leads to a maximum drift length of 350 cm that can be accommodated on the ALTRO memory i.e. much longer than the total length of the LP TPC of 60 cm.

In order to test recent technologies for gas amplification a new charge sensitive premp-shaper (PCA16) has been developed. The programmable PCA16 chip has, as the name indicates, 16 channels and offers different choices with respect to shaping time (30, 60, 90, 129 ns), gain (12, 15, 19, 27 mV/fC), decay time and signal polarity. The new analogue chip required modifications to the Front End Card (FEC), compared to its original design for the ALICE TPC. These are mainly related to the programmability of the PCA16 chip which is done remotely. Data for setting the parameter values are downloaded to the board controller FPGA on the FEC via the data bus on the back plane. An 8-bit shift register delivers the digital input to set the peaking time, the gain, the polarity (common to the 8 PCA16 chips on a FEC) and it also provides a possibility to bypass the shaping function. An octal Digital to Analogue Converter (DAC) controls the decay time of the preamplifier.

Each FEC contains 128 channels i.e. 8 PCA16 chips and 8 ALTRO chips are mounted on each board. They are connected to the pad board on the TPC via flexible 30 cm long Kapton cables. A Readout Control Unit (RCU) governs the readout of the data via the back plane to which a maximum of 32 FEC's can be inserted. Data are sent via an optical cable to a Detector Read-Out Receiver Card (DRORC), which is placed in the Data Acquisition (DAQ) PC. The DAQ software uses the ALICE drivers and libraries for communication between the DRORC and the front end electronics via the optical link. At the reception of the trigger the ALTRO start storing digitised information in the event buffer, up to a predefined number of samples. The RCU reads the ALTRO event buffer and sends the data on the optical link to the DRORC, which stores it in the memory of the readout computer. The events are stored on a local disk whereas a fraction of the events are sent to a monitoring program.

The electronics pedestal and noise levels for all readout channels have been measured both initially as well as on a regular basis during the data taking periods, mainly for pedestal subtraction and to check whether there were corrupted channels. The front end electronics has shown excellent noise performance, with a noise level of 0.5 ADC count corresponding to an equivalent noise of 260 electrons for the longest shaping time and the lowest gain (12 mV/fC), where the noise includes random noise, coherent clock noise and long term variations on the scale of seconds. If the gain is increased to the highest value (27 mV/fC) the noise level increases to typically 1 ADC count, which corresponds to the equivalent noise of 231 electrons at this gain.

Furthermore, the design of a compact version of the ALTRO electronics was completed [21-22]. All components were integrated into a compact 16-channel chip, S-ALTRO-16, and the ASIC was submitted in summer 2010. Delivery of the chip took place beginning of 2011 and at the time of writing the chip is under test.

An alternative approach to the readout electronics was pursued in the form of a TDC based system [23], fully plug-compatible with the ALTRO system and able to operate with the same small pad sizes of order 1x5 mm$^2$ as for the previous one. A total of 640 channels was produced and tested with GEM modules. The operation was found to be difficult due to the low gas gain. For the time being, this option will not be pursued further.

For the Micromegas modules about 1800 channels of the AFTER electronics [24], developed for the TPC of the T2K experiment at JPARC, was used. A more compact version of this electronics has recently been developed and will be used in forthcoming tests.

## 1.13 Test Results

Analyses of data from two different test beam measurements, with the TPC equipped with a two layer GEM system [25], has been performed [26]. Inhomogeneities of the electric field, close to the GEM boarders, caused distortions of the recorded tracks. These distortions were corrected for by using the Millepede program package [27]. During the run the beam was kept fixed and the impact position to the TPC was the same within the beam spread. Millipede makes a least square fit of all tracks simultaneously. After the corrections have been applied the residuals line up around zero.

In Figure 13 (left) the measured space resolution in the direction of the electron drift (z-direction) is plotted as a function of the drift distance for different shaping times. It is shown that the resolution improves with decreasing shaping time until it reaches 60 ns. For smaller shaping times (30 ns) it gets worse, the reason for which is the sampling rate of 20 MHz, which doesn't give enough samples on the rising pulse to determine the front edge accurately. Extrapolating the measurement at 60 ns shaping time to the full drift length of the final ILD TPC (2.15 m) gives a resolution of $\sigma_z(2150) = 446 \pm 9$ μm meeting the final goal of 500 μm. Figure 13 (right) shows the space resolution versus drift distance in the bend plane. A fit to the data points results in an intrinsic resolution $\sigma_y(0) = 59.1 \pm 0.4$ μm. The measurement was performed at a magnetic field of 1 T whereas the magnetic field foreseen for the ILD TPC will be 3.5 T. Theoretical calculation have shown that the space resolution as a function of drift distance at this field is essentially flat so that the result obtained is in accordance with the goal of ≤ 100 μm. Similar results on the resolution in the transverse plane ($\sigma_y$) were obtained with Micromegas equipped with the AFTER electronics.

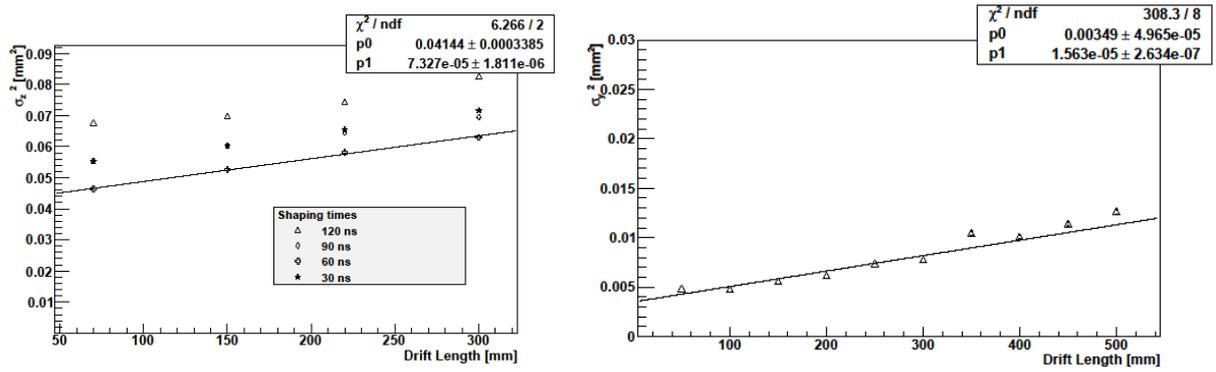

Figure 13: Measured space resolution in the z-direction for different drift distances and shaping times (left) and in the bending plane (right).

## 1.14 Further infrastructure

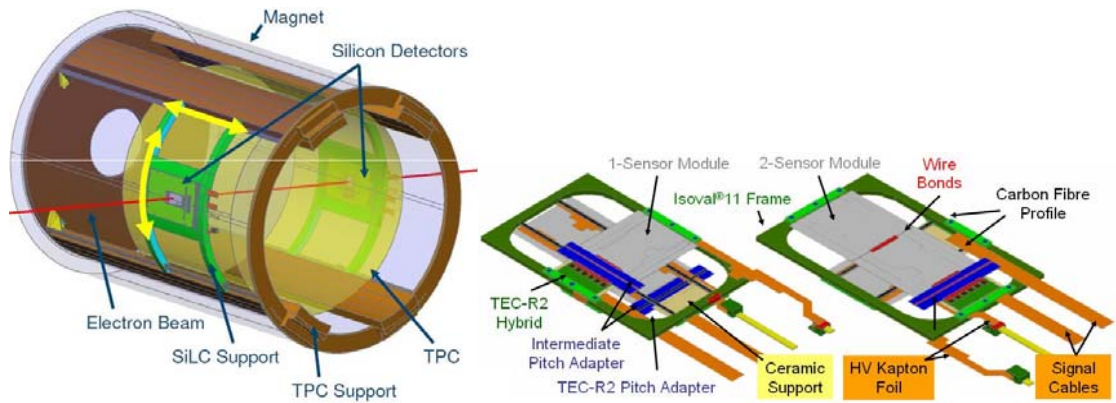

Figure 14: Silicon detectors mounted on a support structure around the LP TPC (left) and a zoom in on the SI-modules (right).

**Silicon Envelope**

In order to have precise external reference points with respect to the tracks within the TPC, a set of highly accurate Silicon strip modules has been deployed on the outside barrel surface of the TPC. These detectors offer a position accuracy of ~ 15 µm in RΦ as well as along the drift axis of the TPC. First beam tests were performed with the main goal to test alignment issues. A schematic view of the setup is shown in Figure 14.

**Supporting Devices**

A set of supporting devices has been installed in order to obtain flexibility in the test beam setup. The TPC itself has to be inserted into the bore of the PCMAG and needs to be moved longitudinally in order to "make use" of the magnetic field (in-) homogeneity. As the inner wall structure of PCMAG is rather thin (1 mm), the TPC had to be supported independently. This has been achieved by a rail system, which is implemented in a cylinder and the cylinder is attached to the frame of PCMAG. In order to make use of the full drift volume of the TPC, one needs to move the TPC with respect to the fixed beam position. The magnet has therefore been placed on a movable stage that allows vertical and horizontal displacements as well as rotation in the horizontal plane of the complete system of PCMAG and TPC (Figure 15). The movements are controlled by a set of step motors and the positions are measured by linear glass scales.

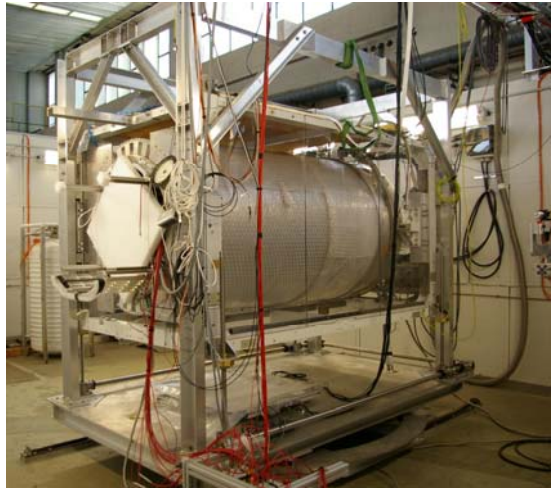

Figure 15: Movable stage around the magnet PCMAG in DESY's test beam area T24/1.

**Laser Calibration System**

The LP cathode surface is made of copper-plated aluminium. A precision pattern of small dots and narrow strips of exposed aluminium was formed by machining away the copper. UV light of 266 nm wavelength from a pulsed laser is used to produce photoelectrons from the aluminium, but not from the copper, giving a well-defined pattern of electron sources that drift to the anode. Comparison of the reconstructed positions of the corresponding charge clusters with the positions of the aluminium dots and strips allows to make distortion studies and to monitor precisely the drift velocity from the arrival times of the charges. A first test run of this system was performed with a Micromegas module mounted on the LP end plate [28].

## 1.15 SiTPC

The SiTPC task aimed to do the necessary R&D for the construction of a precision diagnostic device to measure the electron cloud arriving at the readout plane of a TPC with unprecedented accuracy, both spatially and time resolved. This would be achieved by equipping a TPC like detector with highly pixelised CMOS amplifiers and digitisation ASICs as replacement for the conventional pad plane. Such readout systems could either be used for diagnostic purposes in the Large Prototype TPC system, or be implemented in one or more dedicated (small) monitoring TPCs. It was to be expected that successful construction and results of such devices would be of interest to all groups involved in the development of a TPC as main tracker in an experiment at the ILC and in general to a reviving gaseous tracking detector community as a whole.

In such detectors the charge is directly collected at the input gate of a charge-sensitive amplifier attached to each pixel of the ASIC. This allows unprecedented pad sizes of few $10 \times 10$ μm$^2$. Prior to the start of the EUDET project, first small scale tests using Micromegas and GEMs as gas multipliers and the Medipix2 readout chip as charge-sensitive device were successful [29 - 30]. However, the Medipix2 chip did not offer the possibility to measure the arrival time of the charges at the pixel input gates. One of the main goals of the project was therefore the development of a modified version of the Medipix2 chip, named Timepix, capable of recording the arrival time of the signals. Once the Timepix chip became available, it has first been implemented in small, single-chip detection systems, using both GEMs and Micromegas as gas amplification devices. The final goal was the implementation of a (small) number of Timepix chips into a diagnostic endplate module for the Large Prototype TPC. This was realized with a module consisting of 2x4 Timepix chips and a 3-fold GEM structure for the gas amplification followed by a module consisting of 8 Timepix chips with an integrated Micromegas-like structure (Ingrid) on top of each of the chips in collaboration with University of Twente, Netherlands [31-32].

The development of these new detector technologies would be accompanied by the development of a simulation framework for these systems and of a data acquisition system that could be integrated in the overall DAQ system of the EUDET test infrastructure.

**Development and First Tests of the Timepix Chip**

The main aim of the SiTPC task during the first year of the EUDET program was the development of a first version of the Timepix chip, based on the existing Medipix2 chip [33] that was developed mainly for medical X-ray imaging purposes.

The Timepix chip [34] allows the measurement of the third coordinate, given by the arrival time of the track hits on the pixels. An external clock is distributed to all pixels and the already existing pixel counter starts being incremented, from the moment a 'hit' arrives until the end of the system shutter (common-stop mode). Backward compatibility with the Medipix2 chip was mostly achieved in the new design. Each pixel of the 256x256 array can be selected to operate either in the original 'Medipix' hit counting mode, the 'Timepix' (arrival time) mode or a 'time-over-threshold' mode, providing a measurement of the total charge received on the input of the pixel pre-amplifier.

The layout of the pixel electronics is shown in Figure 16 (left). Initial electronic characterization tests of the chip showed that it behaved according to specification, with a satisfactory yield of good chips of 70-80%. A first beam test was done with a single Timepix chip collecting signals from a triple-GEM gas amplification structure in the DESY T24 electron test beam at 5 GeV. Results, and also from an earlier similar test with a Medipix2 chip for the readout, can be found in [35-36]. Initial tests of the Timepix with a single Micromegas grid as gas amplifier were done, using radioactive sources and cosmic muon tracks. Many detailed studies on single chip detector systems followed. A second production run for the Timepix chip has been performed, with a similar yield of good chips.

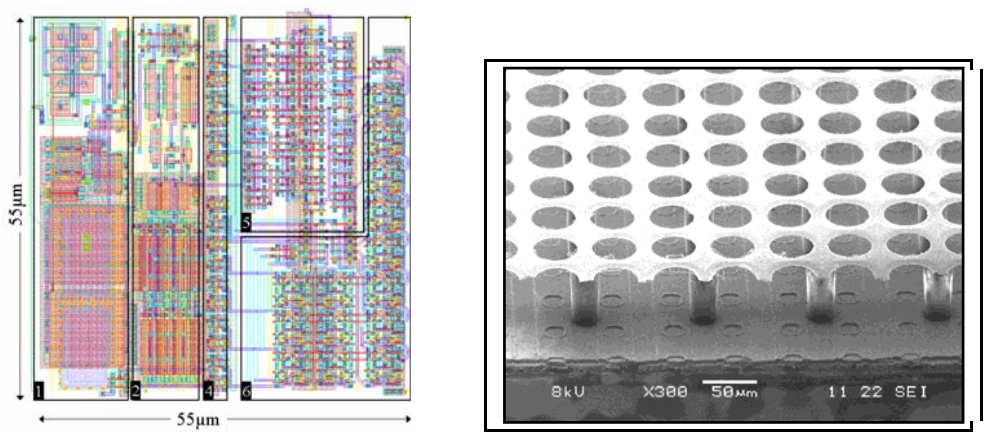

Figure 16: Layout of one pixel of a Timepix chip (left). SEM picture of an Ingrid on top of a Timepix chip covered with protection layer (right).

**Development of Integrated Grids with Discharge Protection**

Initial tests using a Micromegas as gas amplifier showed the appearance of Moire patterns in the response of the device, due to a small mismatch between the pitch of the holes of the Micromegas grid and the pixel pitch of the Timepix, combined with a small misalignment between these two components. Development work for the fabrication of an integrated grid (Ingrid), shown in Figure 16 (right), on top of the Timepix chip using wafer post-processing techniques (on single chips) was also started. Due to (frequent) discharges several of the Medipix2 and Timepix chips were destroyed or sometimes one or more of the control DACs of the chips were damaged. Two possible solutions were pursued: i) the application of a highly-resistive hydrogenated amorphous Silicon layer of 15-20 μm thickness and ii) the production of a double integrated Micromegas like grid, which allows to operate the device with a much lower electric field immediately above the chip.

Both of these proposed solutions proved to be successful. Timepix chips covered with a 15 μm protection layer stayed alive, even after prolonged operation with discharges being provoked by α-particles from a small admixture of Radon in the detector gas flow system [37]. Also layers of $Si_3N_4$ with a resistivity of $\sim 10^{11}$ Ω·cm were applied as protection layer; a thickness of 8 μm proved to be sufficient. A double integrated grid (Twingrid) was successfully produced and tested under a variety of settings for the voltage on the upper and lower grid [38]. Given the successful protection of single-Ingrid devices, the latter Twingrid devices have not been investigated further for the time being.

To increase the mechanical robustness of the Ingrid devices, a GEM like structure was also applied directly on the Timepix chip covered with protection layer. Also this device operated successfully, but with a somewhat reduced gas gain (about a factor two) compared to an Ingrid at the same grid voltage [39]. The first trials of producing such GEMgrids on full wafer scale, were successful.

**Production and Tests of Multi-Chip Systems for a TPC Large Prototype Endplate Module**

The first multi-chip module (two boards of four Timepix chips) was produced and successfully tested at the Large Prototype TPC infrastructure using the DESY T24 test beam. This also involved the development of a readout scheme, consisting of two Windows PC's, each controlling a Medipix Universal Read-Out System (Muros) interface for the control and readout of four Timepix chips, synchronized by a Trigger Logic Unit (TLU, see section 3.2.1) and communicating via TCP/IP with a third Linux PC running the EUDAQ data acquisition system (see section 3.3). The data analysis was done within the Marlin-TPC software framework, which was expanded to allow handling of data recorded with the Timepix ASICs.

The second module consists of a single board with 8 Timepix+Ingrid sensors, being controlled by a single Muros. The detector could not be operated at the nominal grid voltage using T2K gas, due to the appearance of some discharges on two of the Ingrids, However, using a He-i$C_4H_{10}$ (80/20) gas mixture the module was successfully tested in the Large Prototype TPC inside the 1T PCMAG magnetic field at the DESY test beam, with a good single electron efficiency of about 90%.

So far, one 'stand alone' detector with four Ingrids was produced. It was shortly tested, both in lab environment and at the DESY test beams, although in the latter tests some data acquisition instabilities were observed, which are still under investigation.

Figure 17 and Figure 18 show pictures of the first two 8-chip endplate modules, together with examples of some recorded beam track images by the two respective systems.

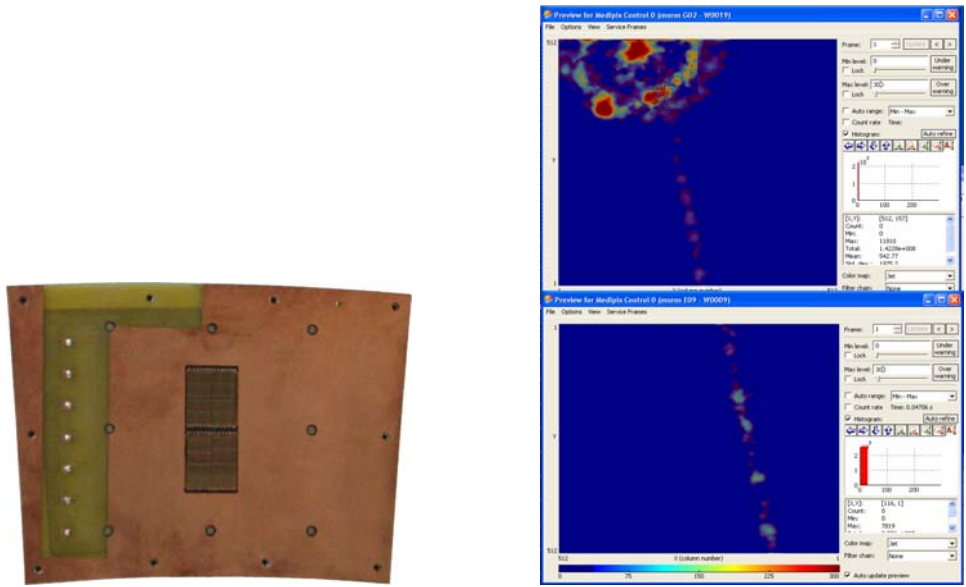

Figure 17: Picture of the readout module consisting of 2x4 Timepix chips; the triple-GEM stack comes on top of this module (left). Example of a beam track recorded by this module (right), showing the individual charge clusters and a delta-ray in the top-left corner

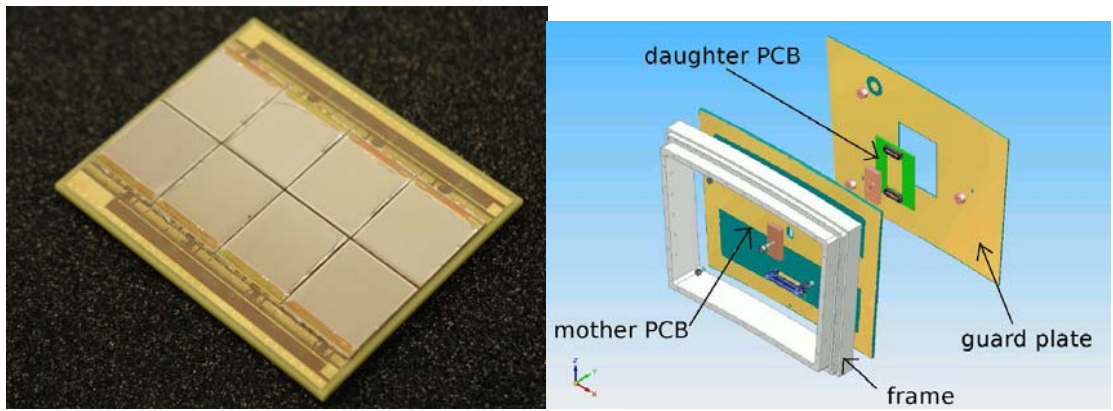

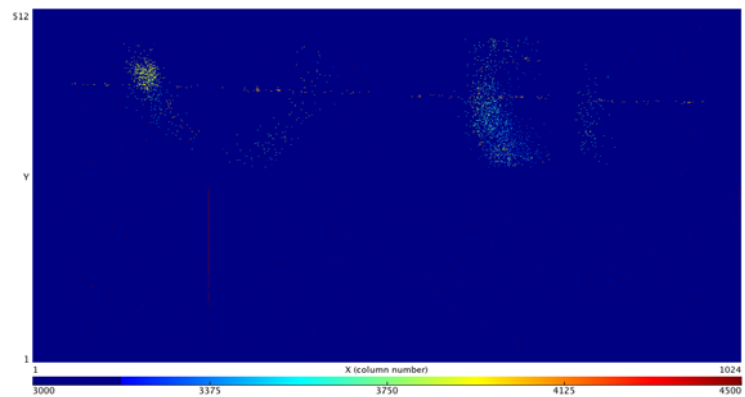

Figure 18: Picture of the daughter board with the 8 Timepix/Ingrid sensors and schematic drawing of the full endplate module (top). Image of 5 GeV electron beam track with two curling delta-rays (bottom).

## 4.8 Silicon Strip Tracking (SiTRA)

The goal of SiTRA was to deliver Silicon tracking test infrastructure and use it for in test beam measurements in transnational activities.

**Building the Silicon Test Beam Infrastructures (SiTI)**

Building a Silicon Test beam Infrastructure (SiTI) implies to construct silicon modules and prototypes, front end electronics to process the signals, a Faraday cage with possibly a cooling system to host the prototypes, an alignment system, a 3D table to allow moving the Silicon prototypes with respect to the beam (IR laser, radioactive source or test beams), an overall DAQ system including both the hardware and the associated software pieces and also the analysis package.

SiTRA benefited of new Silicon strip sensors made by Hamamatsu Photonics K.K. (HPK), which were used to build the modules and prototypes for EUDET. These are large 6" micro-strip sensors, thinner and with smaller pitch than those used by the present state of the art LHC experiments. They include a version with high transmittance used for building the alignment system. Figure 19 shows various silicon modules built (left), and prototypes made of stacked modules (right).

The modules in Figure 19 (right) are hosted in a Faraday cage that can possibly include gas cooling if required because of the power dissipation due to the front end electronics.

The front end electronics is another major ingredient of a SiTI. A full readout chain integrated in a single chip was developed [40]. The SiTRA baseline readout chain included a preamplifier and shaper, a sparsification allowing a trigger decision on analogue sum of 3 neighboring channels, an 8-bit deep analogue sampling plus an 8-bit deep event buffer, an on-chip digitization with a 8-bit ADC and a DAC based calibration.

A version beyond the base line is being developed in a second step. It will include full digital handling of the chip running operation and calibration and slow control management. The corresponding ASIC, labeled SiTR_130 is produced in deep sub micron 130 nm CMOS UMC technology [41]. Figure 20 shows the design, layout of this new Front-End Electronics (FEE) ASIC.

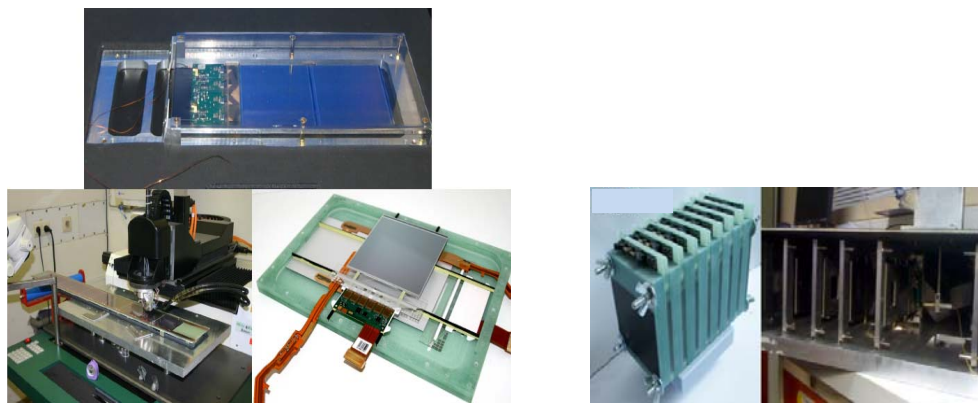

Figure 19: Various silicon modules built within the EUDET project (left); Two examples of silicon stack prototypes for SiTRA (right).

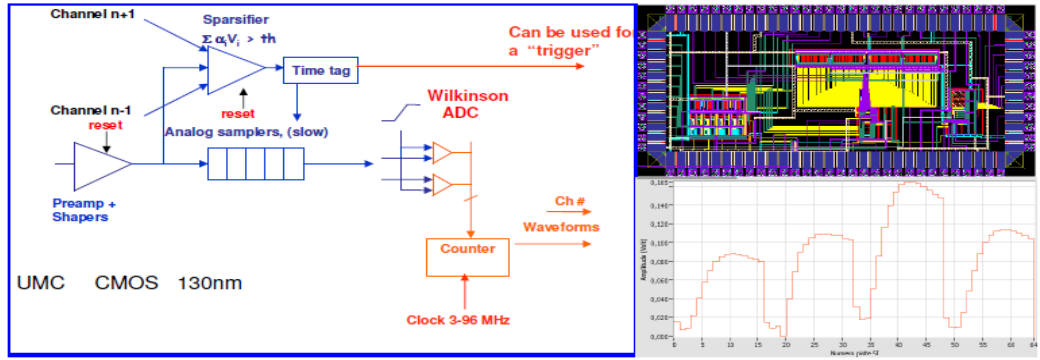

Figure 20: Schema, layout and 4 pulse height distributions reconstructed by analogue sampling of the SiTR_130 front end electronics ASIC

It also shows the pulse heights for 4 channels as reconstructed by the SiTR analogue sampling. This ASIC was delivered, equipped a Silicon prototype at the SPS test beam and gave a signal to noise ratio (S/N) of 23.4 over a microstrip length of about 20 cm. These performances were better than those of the reference VA1' ASIC, produced by the company IDEAS (now GAMMAMEDICA) [42], operating under the same realistic test beam conditions at SPS-CERN.

New versions are developed towards the BBV goals. A first version with 88 channels in CMOS 130nm UMC comprising the full mixed mode architecture was developed and submitted to foundry. The analogue part has been successfully tested. A new version is underway in IBM 130nm with improved front end and ADC blocks and two options, namely a slow and fast version depending the cycle of the collider it will be used at. In addition two reference front end ASIC's were used to equip some silicon prototypes, namely the VA1' ASIC from IDEAS and the APV25 ASIC [43] from the CMS experiment at LHC.

Another important component of SiTI is the laser alignment system [44]. Two options were studied. The baseline option uses the new SiLC HPK strip sensors with some regions of the metalized backplane removed in order to let the IR laser light go through. A complete alignment system was built and successfully tested at a laboratory test bench in Paris with an IR laser and at SPS-CERN. The beam test shows that the S/N ratio for minimum ionizing particles remains constant when scanning regions with or without metalization in the sensors backplane.

The development of higher transmittance sensors is based on detailed simulations, followed by the development of a new sensor technology [45]. As a result, prototyped sensors with a transmittance of 50% were produced on 4" wafers by Institute of Micro-Electronics at Barcelona-National Center of Micro-Electronics (IMB-CNM) (Figure 21), a substantial improvement with respect to the HPK sensors without metallization which had about 20% transmittance. This latter figure is the usual level of transmittance for this type of alignment sensors. The next goal is to transfer to industry the production of these new sensors.

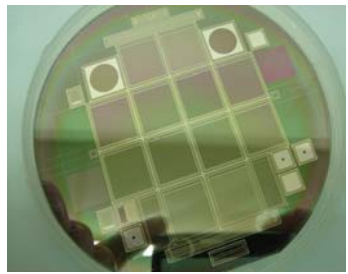

Figure 21: IMB-CNM 4'' wafer with higher transmittance sensors

To complete the mechanical SiTI set up, 3D fully automated tables with LabView based control were built both for the laboratory test bench and for the test beams. They are suitable for testing silicon prototypes (DUT) made of strips or pixels. The DUT can be moved and rotated with respect to the beam line. The 3D Table is built in modular way so that it can accommodate different DUT, with or without alignment telescopes. Five motors are controlled remotely via RS232 to set positions and angles via a LabView application.

To complete the overall silicon test infrastructure, a DAQ system comprising both the hardware and online software were developed. The DAQ system used in the SiTI was based on the existing APV25 [46] related DAQ system developed for CMS [47]. A totally new DAQ system was developed to test the new front end electronics ASIC SiTR and using the VA1' from IDEAS as reference front end chip for comparison. Figure 22 sketches the DAQ hardware chain from the front end boards equipped with VA1' or SiTR chips (for the case of VA1' an additional daughter board with ADC is needed). The digitized signal is then sent to an FPGA that handles the overall data acquisition through a serial USB link to the data taking computer system. On-line DAQ software based on VHDL, C++ and ROOT has been developed. This dedicated DAQ is easily interfaced to other DAQ systems in combined test beams.

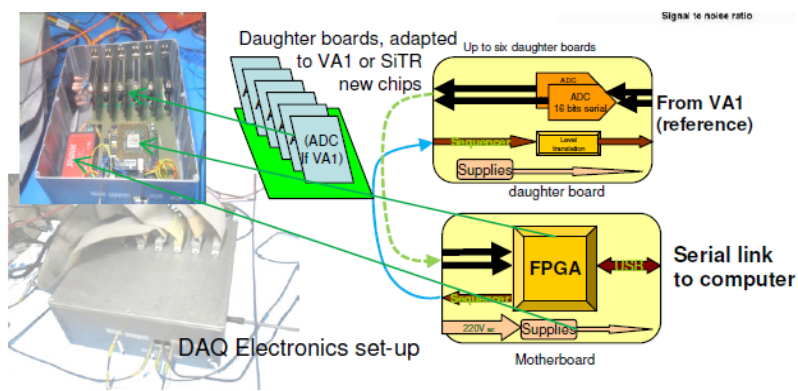

Figure 22: DAQ hardware chain to read out and fully handle the data taking of silicon prototypes with different front end electronics

**Test Beam Campaigns**

Several SiTI systems were built within the EUDET project, which allowed a series of test beam measurements with valuable results. Figure 23 shows a photograph of the first combined test beam set up with a SiTRA test infrastructure comprising three Silicon modules and the EUDET telescope at the SPS-CERN. They are made of 3 HPK micro-strip sensors read out by the new SiTR_130 FEE or by the reference VA1' chips.

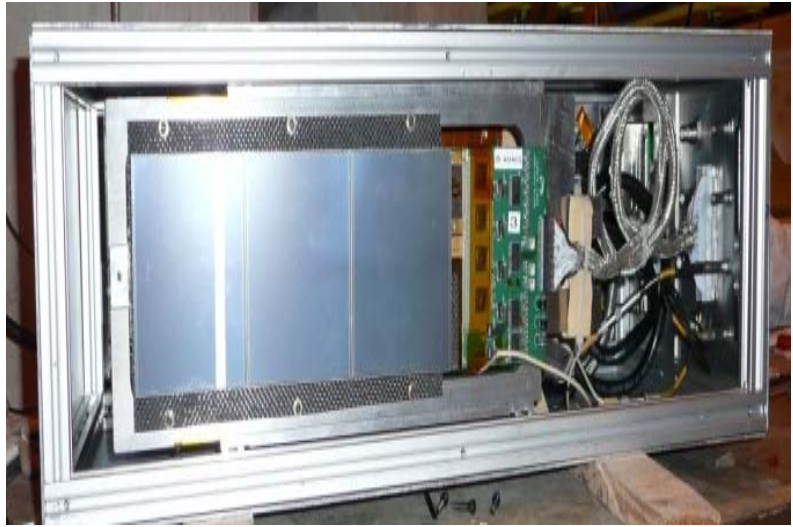

Figure 23: First combined test with SiTRA and the EUDET telescope at the CERN SPS, with new SiTR based readout.

Figure 24 shows another combined test beam set-up with the EUDET telescope performed by the SiTI, containing a set of 8 Silicon strip modules made with HPK sensors (Figure 19), read out with APV25 chips. New inline pitch technologies and new sensors, in particular new 6" Double Side Silicon Detector (DSSD), were tested with interesting results. Figure 24 on the right shows the standalone multipurpose SiTI, that tested the HPK based alignment system and new 3D pixels. The SiTI is installed on the 3D table.

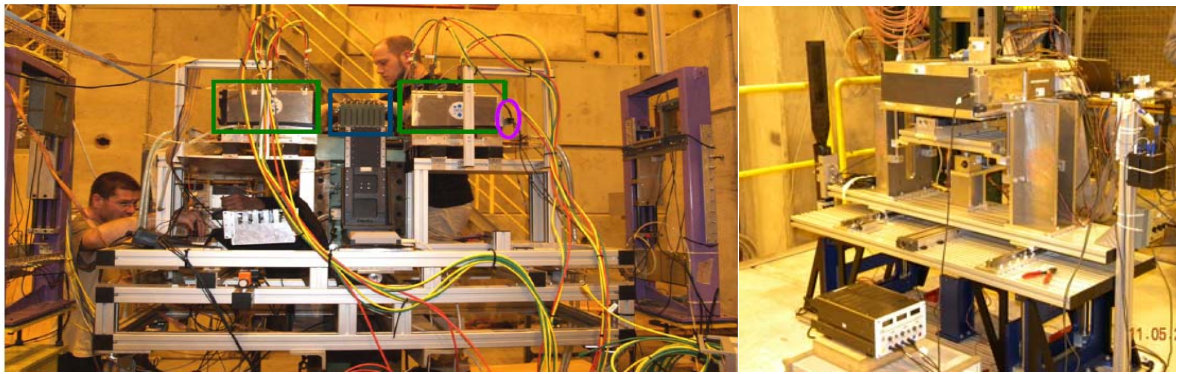

Figure 24: Combined test beam with EUDET telescope and SiTI, based on APV25 readout (left). Standalone and multipurpose SiTI set up on the 3D automated table (right)

A complete chain of analysis programs was developed to analyze the data recorded in the beam tests and the Prague team was instrumental in developing such packages including full event reconstruction, tracking fits and simulations.

**Perspectives for the Future Transnational Activities**

Several test beam campaigns were performed this last year or will be achieved after the completion of the EUDET program with the SiTI. The SiTI were or will be instrumental to test new Silicon sensors based on improving the existing strip technology by studies of new patterns or novel technologies. These test infrastructures are also an important asset to develop the new FEE readout mixed mode ASIC able to fully process the signals from a large number of channels including digitization, and a highly fault tolerant and flexible data handling.

The development of DAQ systems for these test infrastructures is essential for developing the final DAQ of large area silicon tracking system and to test every component.

EUDET was crucial for advancements in many of the important items to build the next generation of large area Silicon trackers. The developed expertise and test systems will extend much beyond the EUDET project.

# Calorimetry

The R&D for calorimetry was organized in five tasks, which prepare infrastructure for research and development on calorimetric sub-detector systems. These comprise calorimeters for electromagnetically (ECAL), hadronically (HCAL) interacting particles and for particles emitted in the forward direction (FCAL), as well as the associated readout systems i.e. front end electronics (FEE) and data acquisition (DAQ). The overall goal of the detector R&D was to explore the use of novel sensor technologies for finely segmented calorimeters, and it proceeds now from proof-of-principle prototypes currently being exposed to test beams towards realistic i.e. scalable detector prototypes. The infrastructure initiative aimed at providing realistic environments (mechanical structures, readout systems, test stands) for the tests of the new technologies being proposed by R&D groups for use in a future ILC detector.

## 1.16 Electromagnetic Calorimeter

The ECAL "EUDET module" (Figure 25) is a large scale module ($160\times30\times20$ cm$^3$) for the ILC, addressing the industrialisation issues as well as the cost and system aspects. The design of the module features slabs that are sled inside a mechanical tungsten/carbon fibre structure that provides alveoli for the slabs (Figure 26). A single slab consists of a carbon/tungsten H shaped board ($150\times18\times2.1$ cm$^3$ and $150\times18\times4.2$ cm$^3$, respectively) which will be equipped with active units on both sides of the tungsten core.

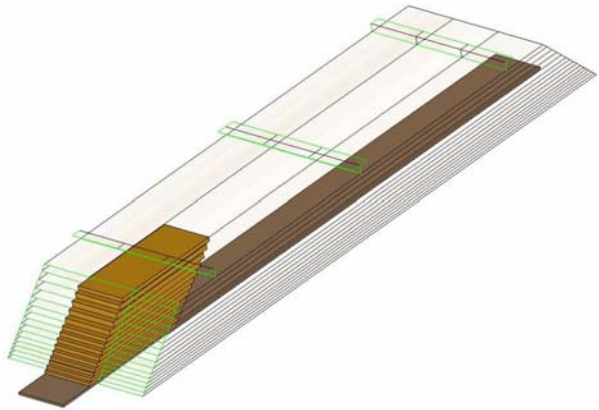

Figure 25: Conceptual drawing of the EUDET ECAL module, equipped with a few full-length slabs and a projective tower containing an electromagnetic shower

A description of the base line for the construction of the module can be found in [48], which also addresses the technological challenge to construct a compact and highly granular calorimeter. This concerns in particular the integration of the Front End Electronics into the layer structure of the calorimeter, which has to be embedded in slits of the order of 1mm. All elements of the detector represent technology reaching beyond nowadays state-of-the art.

Given the uncertainties on the final Si wafer dimensions it was decided to construct a so-called Demonstrator benefiting thus from already existing material such as tungsten absorber plates.

The H structure for the Demonstrator does have nearly the same dimensions (150x13x4.2 cm$^3$) as for the EUDET Module and the alveolar housing is very similar to the one envisaged for the EUDET Module. The Demonstrator thus allows for the validation of important engineering issues such as the fabrication of large size H-structures and alveolar structures.

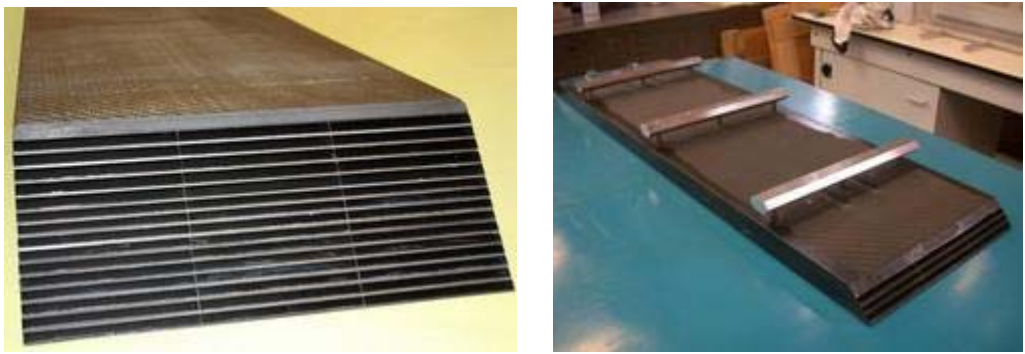

Figure 26: View of the alveolar structure of the EUDET ECAL mechanical prototype

The assembled alveolar structure has been produced and parallel to that, the corresponding parts have been designed and ordered for the EUDET module. An important part of the test program with the Demonstrator will be the study of the heat dissipation by the Very Front End (VFE) electronics embedded into the detector layers and the way how to evacuate the generated heat. The slab for the Demonstrator was equipped with thermal boards, which simulate the heat generation. The Demonstrator allowed for the verification of existing simulations of the heat dissipations. The cooling system to equip the Demonstrator is designed such that it can be scaled up to the needs of the EUDET Module.

A batch of 30 Silicon wafers with the correct dimensions but a different number of readout pads (Figure 27) as finally foreseen for the EUDET Module has been delivered by Hamamatsu (Japan) and tested. Apart from minor problems, the quality of the wafers has been found suited for the employment in the EUDET Module. Wafers for the EUDET Module with 256 pads have been ordered at Hamamatsu and OnSemi (Czech Republic). In parallel studies of how to apply the bias voltage to the Si wafers are ongoing.

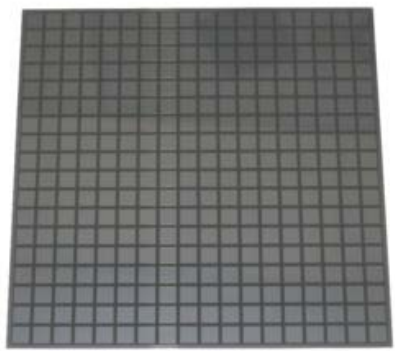

Figure 27: Photograph of the Silicon wafers produced by Hamamatsu. This wafer contains 324 pixels.

The PCB will finally have a thickness of 1.2 mm in order to accommodate 256 channels as read out by four SKIROC chips, which are mounted on it as seen in Figure 28 (left). Studies with a prototype board shown in Figure 28 (right), have revealed problems during the bonding of an ASIC to the board.

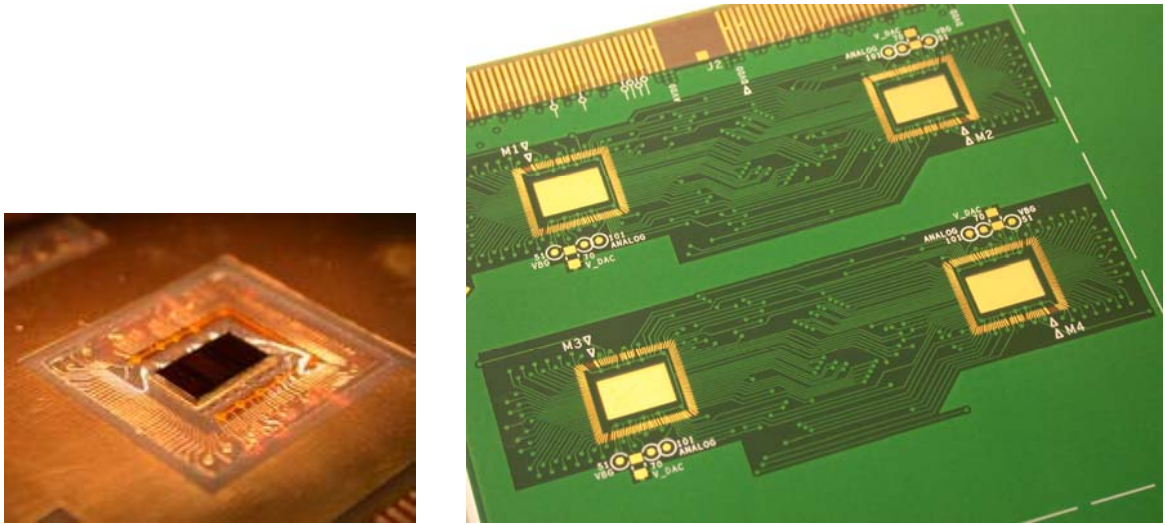

Figure 28: ECAL PCB (ASU) with a chip SPIROC bonded (left) and a Prototype board (right)

## 1.17 Hadronic calorimeter (HCAL)

**Mechanical Integration of the Hadronic Calorimeter**

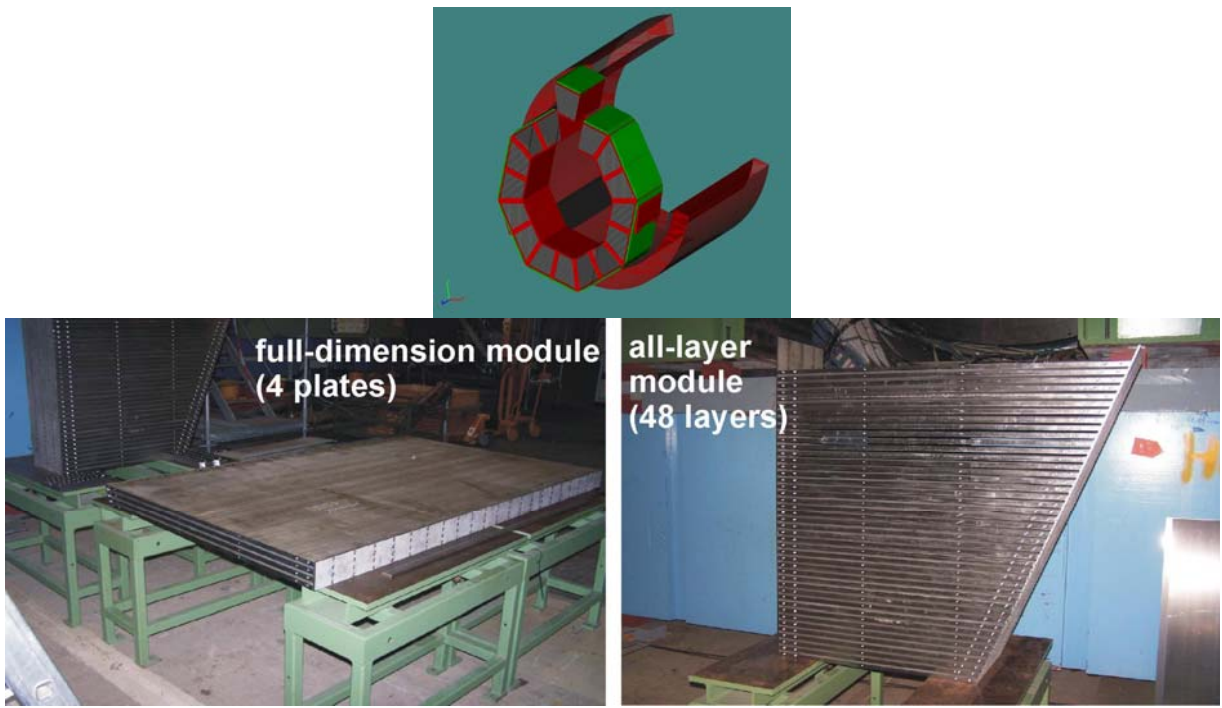

Figure 29: Conceptual view of one half of the HCAL barrel, which is subdivided into 16 sectors (top) and surrounded by the magnet, and realized mechanical test modules for a horizontal set-up with the full half-barrel length of 2.20m (bottom left), and for the vertical setup with 48 layers (bottom right).

Two mechanic structure prototypes have been realized during the course of the EUDET project: A horizontal test structure (Figure 29, bottom left) to establish the mechanical tolerances over the full area of an HCAL half-barrel segment. Heat dissipation and installation procedures can be studied in full scale. In addition, two identical vertical test structures (Figure 29, bottom right) have been built to establish the mechanical stability under various orientations and stresses. The vertical test structure can be stacked side by side, to test the mechanical behavior of the module interconnections, or behind each other, to build up test beam structures.

Both test structures can be equipped with the active-layer's electronics in order to test the full-scale performance of the electronics, especially the signal integrity, crosstalk behaviour and grounding strategies. The vertical test structure can be used for testbeam studies of the detector concept in various orientations with respect to the beam.

The optimization (minimization) of the gap between the absorber plates requires tight tolerances concerning the flatness of the absorber plates. The typical flatness of commercially available, rolled steel plates at sizes of 1x2.20 m² is not better than 2-3 mm. A flatness of below 1 mm of machined plates would increase the costs by a factor of 2 to 3. Since cost optimization is important for a realistic design concept, standard rolled plates have been used for the test structures. These have been optimized before assembly in a so called roller-levelling process, by which flatness below 1mm could be achieved at very low cost.

**Electrical Integration**

In Figure 30 the realized prototype electronics, HCAL Base Unit (HBU), is shown. The HBU with a typical size of 36x36 cm² integrates 144 scintillating tiles each with Silicone Photomultipliers (SiPM) together with the front-end electronics and the light calibration system. The analog signals of the SiPMs are read out by four front-end ASICs [49]. In order to cover the full depth of 220 cm of a segment's layer, 6 HBUs are connected together, forming an electrical layer unit within a layer cassette (slab). Further specifications of the HBU can be found in [50].

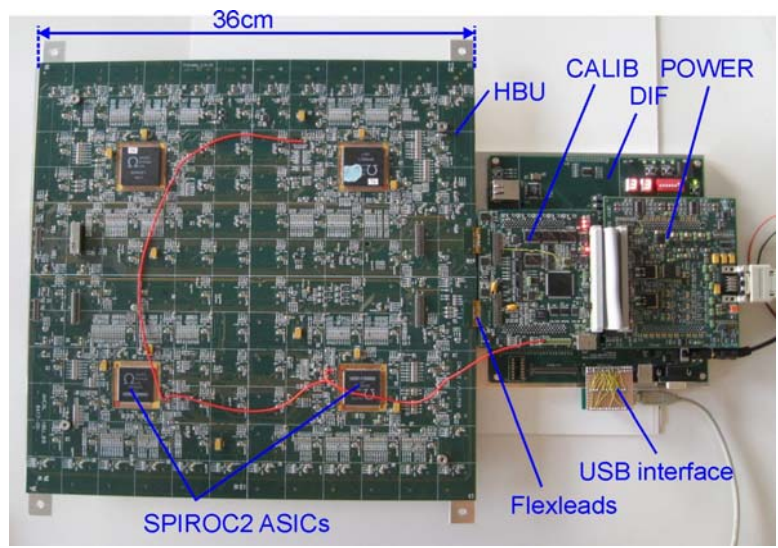

Figure 30: Prototype of the realized inner detector module HBU and the interface modules.

The DAQ interface boards DIF, CALIB and POWER form the interface between the inner detector electronics and the global data acquisition, via the CALICE DAQ module Linked Data Aggregator (LDA) [51].

Two prototypes for the LED light distribution system have been realized, both are based on the usage of blue or ultraviolet LEDs. The first approach is based on only few strong LEDs outside the inner detector, and a light distribution via optical fibers into the detector on top of and across the HBU PCBs [52-53]. The light is coupled through notches in the fiber and holes in the PCB into the tiles. The uniformity of the light radiated by the notches has been measured to be constant within 20%. In the second approach, one LED is assembled per tile (144 pieces per HBU), offering the advantage that no fibers are needed. The LED trigger can be distributed electrically and differentially. Measurements with a test system have shown no crosstalk from the LED trigger to the SiPMs. The channel-to-channel uniformity is under test and will be optimized in the next integration step by a new LED driving circuit and a new type of LEDs.

Both systems have been tested together with the HBU and the interface modules in the DESY testbeam environment within the EUDET Transnational Access Program. Linearity and dynamic range of the approach are under investigation.

**Measurement Results of the Prototype Electronics**

Two HBU prototype systems have been realized and tested in the laboratory for basic characterizations and at the DESY electron testbeam facility. In Figure 31 (left), a typical MIP spectrum is shown, which has been measured with the highest gain of the SPIROC2 front-end ASIC. The single-pixel structure is distinguishable from pedestal up to far behind the most probable value of the MIP. With this setup all tiles have been scanned and the light-yield for a MIP signal was extracted. The average light yield for 26 channels resulted to be 8.04 pixels/MIP with a spread of 0.9 pixels (standard deviation).

The integrated LED calibration system has been tested in both setups. For most of the channels characteristic single-pixel spectra as shown in Figure 31 (right) were obtained. The distances of the pixel-peaks in the histogram are a measure for the respective channel gains and will be used in dedicated LED calibration runs for gain monitoring.

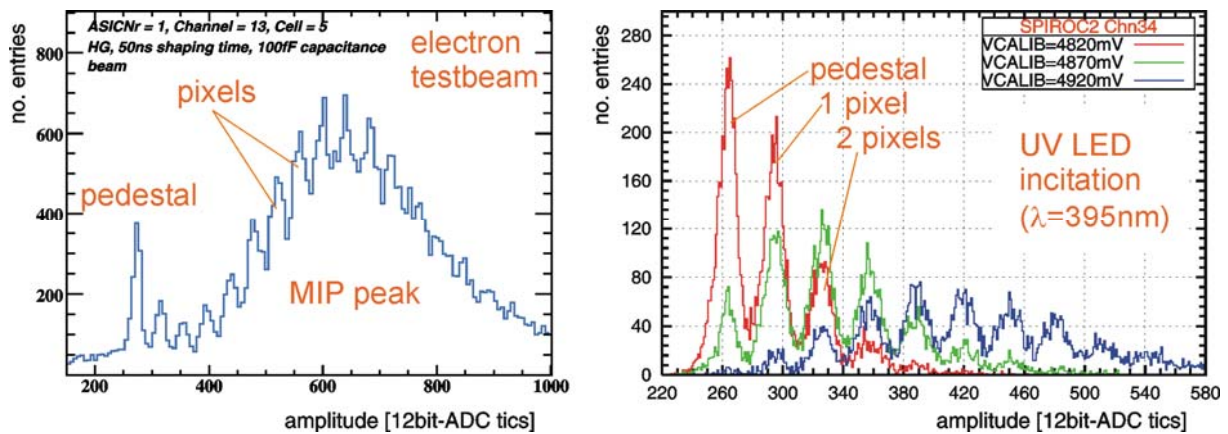

Figure 31: Exemplary results from single-channel analysis with testbeam and the LED calibration system. Single-tile response of one channel (left) and the characteristic single-pixel spectrum of one channel for three different LED-light amplitudes (right).

Since in the ILC experiment no central trigger is provided to the front-end electronics, the SPIROC2 ASICs have self-triggering capability. The ASICs compare the analog input signal after amplification and fast-shaping with a predefined threshold and generate a trigger decision. The threshold can be adjusted by an internal 10-bit DAC, the output voltage of the DAC is set within the SPIROC2 slow-control configuration. In a dedicated analysis, the threshold dependence on the DAC value has been determined for different detector channels. The resulting significant spread among the channels leads to the requirement of defining the trigger threshold individually per channel.

In order to determine the MIP detection efficiency in self-triggering mode, the threshold has been set in a way that the ratio of the number of noise events and the number of MIP events is below $10^{-4}$. With this threshold cut applied, a MIP detection efficiency of 95% for the analyzed channels could be achieved in accordance with system specification.

The self-triggering mode of the system is the most important mode of operation with respect to an application for the ILC and therefore its performance has to be tested carefully.

**Final Integration**

In order to achieve the high integration level as shown in Figure 29, the prototype modules shown in Figure 30 have to be redesigned. The HBU redesign (HBU2) retains its general shape and outer dimensions of 36x36 cm², but the SPIROC ASICs are replaced by 4 ASICs of the newest generation SPIROC2b which are currently under test. The proposal for the final interface module's setup is shown in Figure 32. The commercial FPGA board is replaced by an own development, still based on a SPARTAN3 FPGA in order to be design-consistent with the other CALICE DIF modules. The modules DIF, CALIB2 and POWER2 are realized as mezzanine modules on the so called central-interface board (CIB). All active electronics on the interface boards that might need an exchange due to a failure over time are connected with robust interface connectors to the CIBs (Samtec MEC1-RA type) in order to guarantee an easy replacement. The CIB has a typical width and length of 36 cm and 10 cm, respectively. While the CIB connects to the layer's middle slab, the Side Interface Boards (SIBs) connect to the side-slabs (c.f. Figure 25). The SIBs only contain passive electronics and act similar to a cable. As interconnection between CIB and slab, CIB to SIBs and SIB to HBUs, the flexleads of the physics prototype can be used.

With up to six of the HBU2s and the CIB with the mezzanine interface boards the layer performance in the horizontal mechanical structure can be tested, a tower module with only one HBU connected to the CIB in each of the 48 layers of the vertical mechanical prototype can be realized as well.

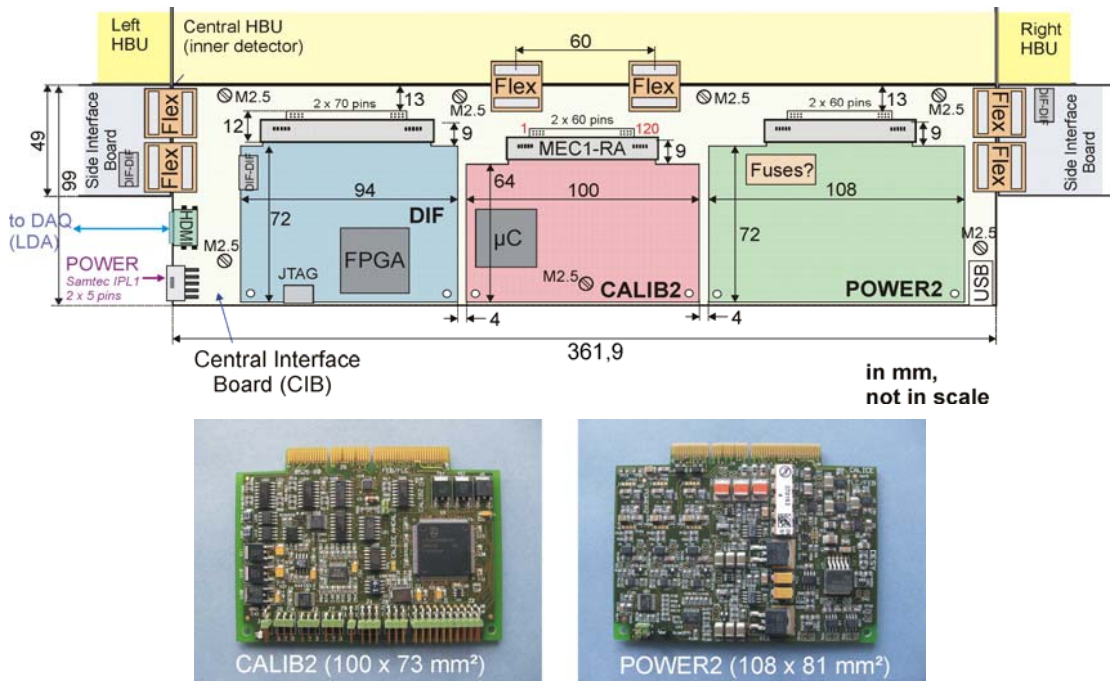

Figure 32: Dimensional sketch of the final integration step for the interface boards (top), and photographs of the already realized modules CALIB2 and POWER2 (bottom).

## 1.18 Forward Calorimeter (FCAL)

The objective of FCAL is the development of the infrastructure to investigate several promising technologies for two special calorimeters, BeamCal and LumiCal, in the very forward region of a future collider detector. These calorimeters must be compact, precisely positioned with a very fast readout. The readout electronics was successfully tested in the laboratories. In particular a redesigned ADC ASIC with several technological improvements was produced and successfully tested in the laboratory. Essential parameters like integral and differential nonlinearities match the requirements. Sensor prototypes made of silicon for LumiCal and GaAs for BeamCal have been tested in the laboratory and in particle beams in previous years using the infrastructure supported by EUDET. These sensors have been assembled with front-end ASICs and investigated first in the laboratory and then tested in a particle beam at DESY. Here results from the test of the ADC ASICs and from the investigation of an assembled sensor plane are reported.

**Read-out Electronics**

The block diagram of the LumiCal readout scheme is shown in Figure 33. Prototypes of front-end ASICs [54] and single channel version of the ADC were already designed, produced and tested in previous years. The main effort was aimed to the following issues:

- Complete the characterization of a single channel 10-bit pipeline ADC
- Design and first prototyping of a multichannel 10-bit pipeline ADC

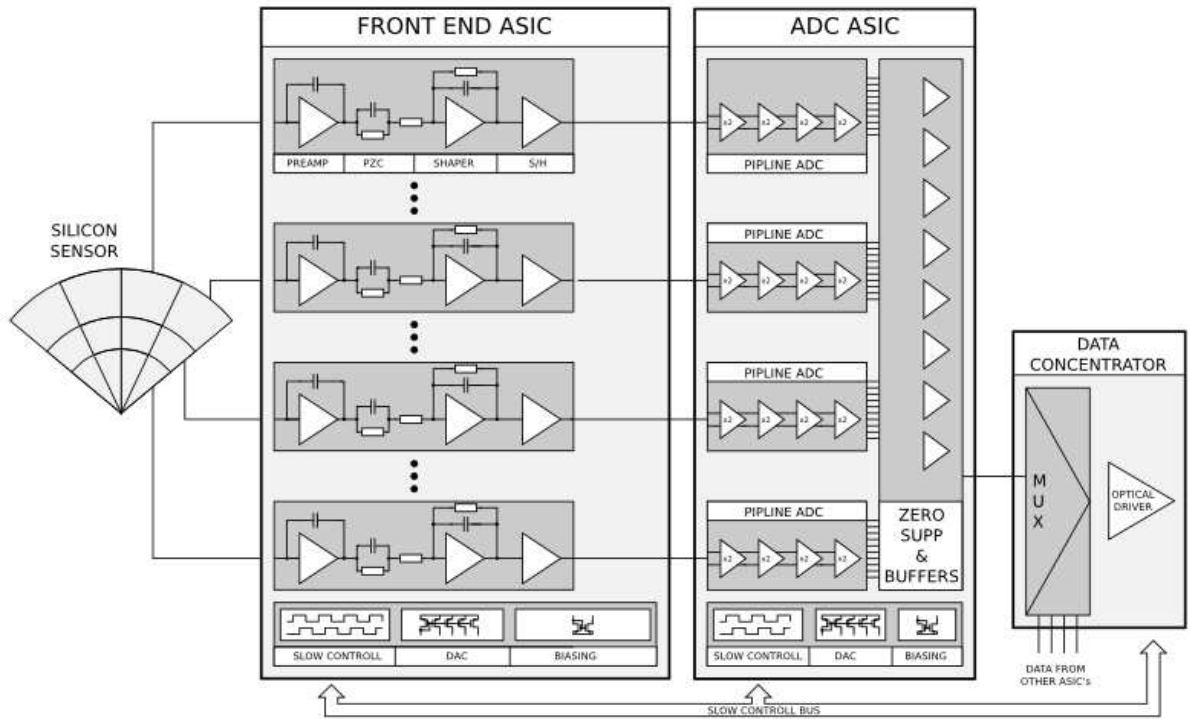

Figure 33: Block diagram of LumiCal readout electronics

The prototypes of a single channel 10-bit ADC were produced in an early stage and partially tested. The later studies were concentrated on:

- The power consumption as a function of the sampling frequency and the supply voltage.
- Characterization of the static performance, i.e., the measurements of integral (INL) and differential (DNL) nonlinearities
- Characterization of ADC dynamic performance, i.e., the measurements of signal to noise ratio (SNHR and SINAD) and harmonic distortions (THD)
- Studies of the fast powering on/off feature.

All parameters mentioned above were measured and found to be in agreement with expectations from simulations. The complete discussion of the results can be found in [55]. The list of the key parameters obtained is shown in Table 1. The possibility of switching off the power lowers drastically the average power consumption in applications with long idle times. For standard ILC beam conditions it allows to decrease the average power consumption down to about 15 μW.

Table 1: ADC measured parameters architecture 10-bit pipeline

| architecture | 10-bit pipeline |
|---|---|
| technology | 0.35 μm CMOS |
| sampling rate range | 1 kS/s–25 MS/s |
| input range | 2 Vpp |
| power consumption | ~0.85 mW/MS/s, 3 Vsupp (~0.6 mW/MS/s, 2.6 Vsupp) |
| area | 0.87 mm$^2$ |
| linearity | INL < 1 LSB, DNL < 0.5 LSB |
| SINAD | 56–58.5 dB |
| ENOB | 9.3 bit |
| time of switching OFF/ON | ~3 μs |

Since the

single channel ADC fulfilled the requirements the design of a multichannel version was done as the next step In the first version 8 channels of the above described single 3 channel version was layout in the prototype ASIC. Because of various settings and control signals needed, a number of new functionalities and technological improvements have been implemented in the multichannel ADC version. The submitted prototype ASIC contains:

- 8 channels of the core pipeline 10-bit ADC.
- Digital multiplexer/serializer providing different modes of operation. In the Serial mode, which is the ILC baseline mode, one data link is used for output of all ADC channels, allowing a maximum ADC sampling frequency up to about 4 MSps. In the parallel mode one data link serves per each channel, allowing a maximum sampling frequency of about 25 MSps. In the test mode a single channel is readout through 10 parallel output lines.
- High speed LVDS drivers (1GHz) with variable power consumption
- Low power DAC control references for various settings of ADC biasing currents and voltages.
- Precise Band-Gap reference source.
- Temperature sensor.

The layout of the prototype ASIC is shown in Figure 34.

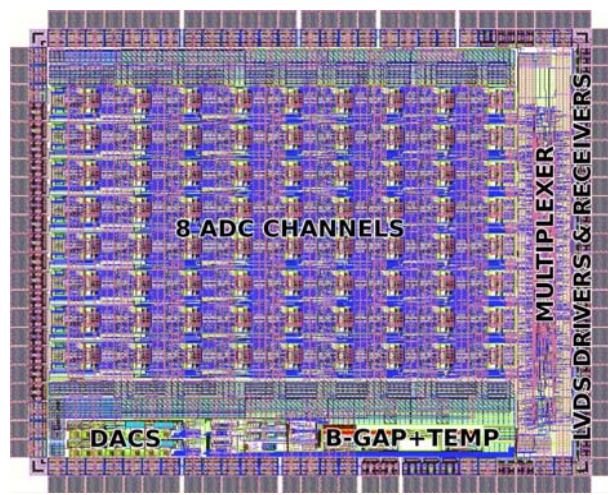

Figure 34: Layout of multichannel ADC

The tests of the multichannel ADC prototype have been started. Preliminary measurements confirm the expected ASIC functionality. In Figure 35 measurements of the FFT spectrum (top) and dynamic parameters (bottom) as a function of the sampling frequency obtained in the test mode are shown. It is seen that the multichannel version works well up to about 45 MHz, which is substantially higher than for previous, single channel version. The measurements are just ongoing and more results will be available soon.

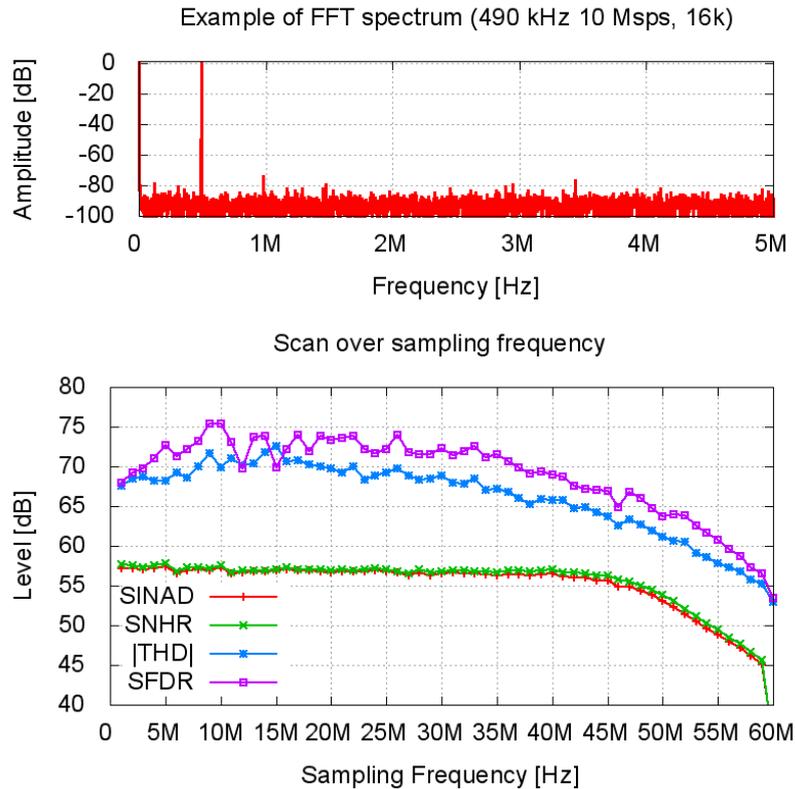

Figure 35: Example of dynamic ADC measurements: FFT spectrum (top), dynamic parameters vs sampling frequency (bottom)

**Assembled Sensor Plane Tests**

Using sensor prototypes made of silicon and GaAs, as shown in Figure 36 for LumiCal and BeamCal, respectively, sensor plane prototypes have been assembled with front-end ASICs. These sensors were supported by a PCB. Groups of 8 strips and pads of several areas of the sensor were interconnected via wire-bonds to 8-channel front-end ASICs. After amplification and shaping signals were driven to V1724 (14-bit, 100 Ms/s) and V1721 (8-bit, 500 Ms/s) Flash-ADCs in a VME Crate. The sensors were positioned in a 4.5 GeV electron beam using a movable x-y table. Beam particles were triggered by three scintillator counters and precisely measured by three planes of silicon strip detectors, two upstream and one downstream of the sensor under test, as shown in Figure 37. The sensor planes were moved in several positions to record data from particles crossing different pads and in particular also from particles crossing the areas between the pads and the edges of the planes. Within a few days several million trigger have been collected. In a few runs tungsten absorber plates of different thickness were positioned in front of the silicon sensor plane in order to measure particle showers.

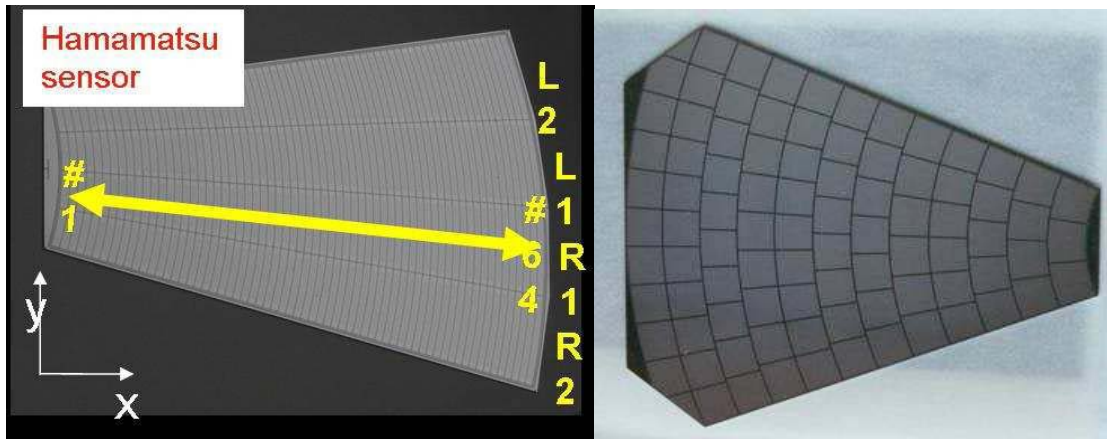

Figure 36: Prototype of silicon sensor (left) and a GaAs sensor (right)

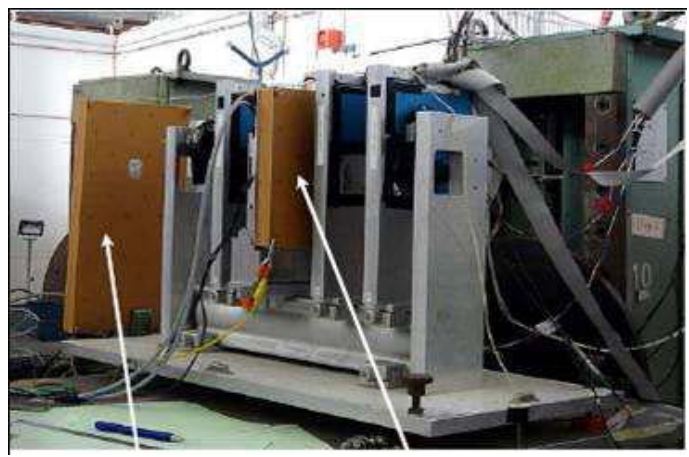

Figure 37: The setup used for the beam-test. The three planes of silicon strip detectors are inside the aluminum frames. The sensor plane under test is mounted inside the copper-colored PCB box between the second and third aluminum frame.

**Preliminary Results**

Two examples of signal spectra from single beam particles are shown in Figure 38. The spectra follow nicely the Landau distribution of energy loss. The signal-to-noise values are around 18 for LumiCal. The channel-by-channel differences in the amplification are below 1%. Also the cross-talk between channels is less than 1%.

Using the telescope the impact point of the beam-particle on the sensor is predicted. The signal of the pad hit by the particle is measured, and if it was found above the pedestal, the impact point coordinates are fed into two-dimensional histograms, as shown in Figure 39. The structure of the pads on the sensors, indicated by different colors, becomes clearly visible.

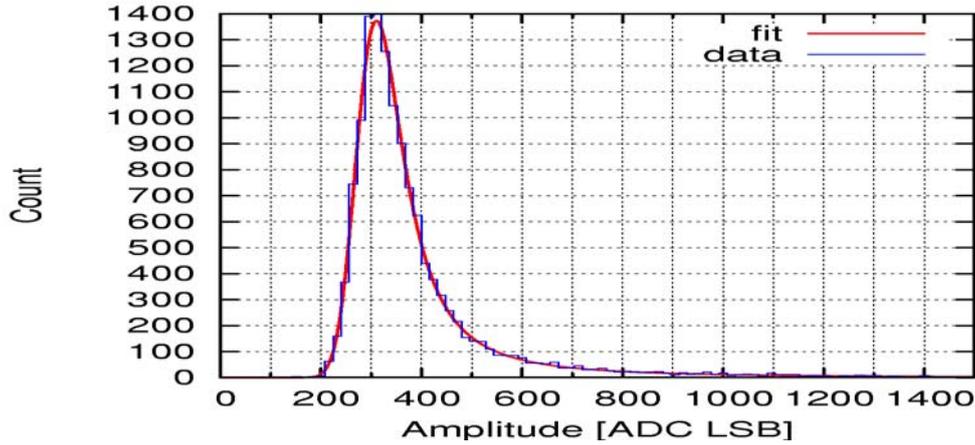

Figure 38: Examples for signal spectra obtained from the sensor plane prototypes of LumiCal

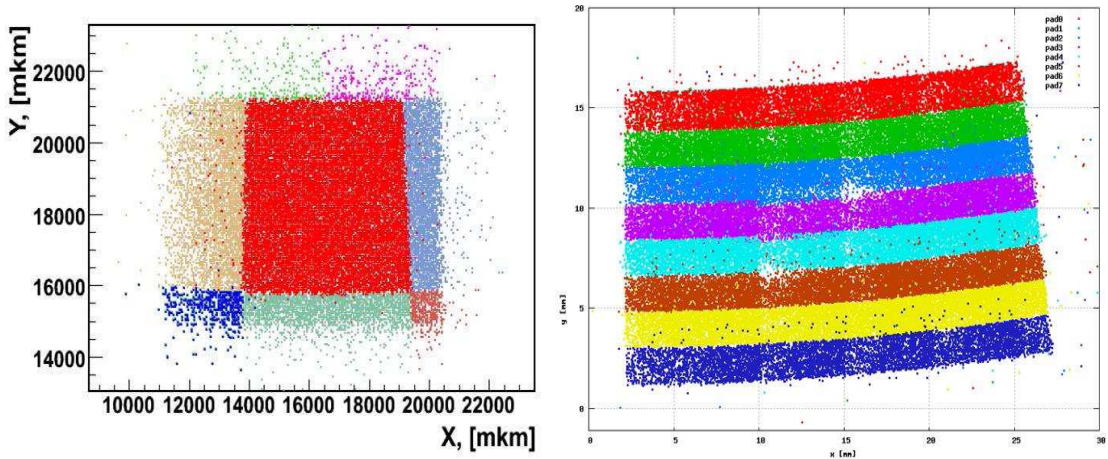

Figure 39: The distribution of impact point of beam particles on the sensors for BeamCal (left) and LumiCal(right). Colors are assigned when the signal on the pad hit is above the pedestal.

Data were also taken with a few radiation length of tungsten in front of the LumiCal sensor plane to record shower particles. Both the distribution of the measured deposited energies and the average deposited energy as a function of the tungsten thickness are well reproduced by Geant4 simulations of the set-up. The test-beam data analysis is still ongoing. More refined results will be available soon.

## 1.19 Front-End Electronics

In the framework of EUDET, a second generation of readout ASICs has been developed to readout the technological prototypes described above. They are based on the 1st generation of chips that were used for CALICE physics prototype for the analog front-end part but add several essential features:

- Auto trigger to reduce the data volume
- Internal digitization to have only digital data outputs
- Integrated readout sequence and common interface to the 2nd generation data acquisition to minimize the number of lines between chips
- Power-pulsing to reduce the power dissipation by a factor 100

Three chips have been designed:

- HARDROC for digital Hadronic Calorimeter (DHCAL), for RPCs or Micromegas chambers. A new ASIC, MICROROC, has also been designed for 1 m$^2$ MICROMEGAS detectors, which require HV sparks robustness for the electronics and very low noise performance to detect signals down to 2 fC.
- SPIROC for analog Hadronic Calorimeter.
- SKIROC for the W-Si Electromagnetic Calorimeter.

These ASICs (except MICROROC) were produced (Figure 40) in a dedicated run of AMS SiGe 0.35 µm technology. 25 wafers, each with 500 HARDROC2, 70 SPIROC2A, 70 SPIROC2B, 70 SKIROC2 have yielded a few thousands of chips that can equip respectively DHCAL, AHCAL and ECAL modules. The chips have been packaged at the I2A Company (USA) in an ultra-flat TQFP package to be embedded inside the detector modules with a minimal thickness.

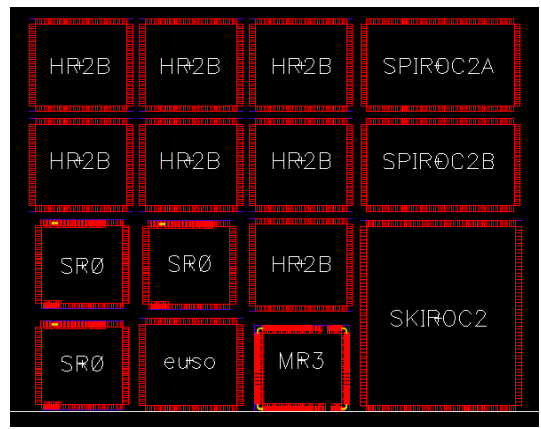

Figure 40: Reticle of the production run

**HARDROC ASIC**

HARDROC readout is a semi-digital readout with three thresholds (2 bits readout) which allows both good tracking and coarse energy measurement, and also integrates on chip data storage. The chip integrates 64 channels of fast low impedance current preamplifier with 6 bits variable gain (tuneable between 0 and 2), followed by a fast shaper (15 ns) and low offset discriminators. The discriminators feed a 128-deep digital memory to store the 2*64 discriminator outputs and bunch crossing identification coded over 24 bits counter. Each is then readout sequentially during the readout period.

A first version was fabricated in AMS SiGe 0.35 µm technology and met design specifications. A second version was produced to fit in a smaller low-height package (TWFP160) which necessitated changing the double-row bonding pad ring into a single row, rerouting all the inputs and removing many pads (Figure 42). A possibility for a third threshold was added at that time, also separating more widely the three thresholds (typically 0.1-1-10 pC) and the "off" power dissipation was brought down to a few µW for the whole chip.

The trigger efficiency allows the MIPs for RPCs to be discriminated with 100 fC threshold (10 fC for Micromegas) with a noise of 1 fC, shown in Figure 41 (left). The power pulsing scheme has also been validated, shown in Figure 41 (right), where 25 µs are required to start up the chip so that it can trigger on a 10 fC input signal. Finally the readout scheme, which is common to all the chips, has been validated on the large square-meter board, built as a scalable technological prototype of DHCAL that is read-out by the side.

This HARDROC chip is the first one on which large scale power-pulsing has been tested at system level, allowing a power reduction by a factor 100 while keeping the detector efficiency above 95%.

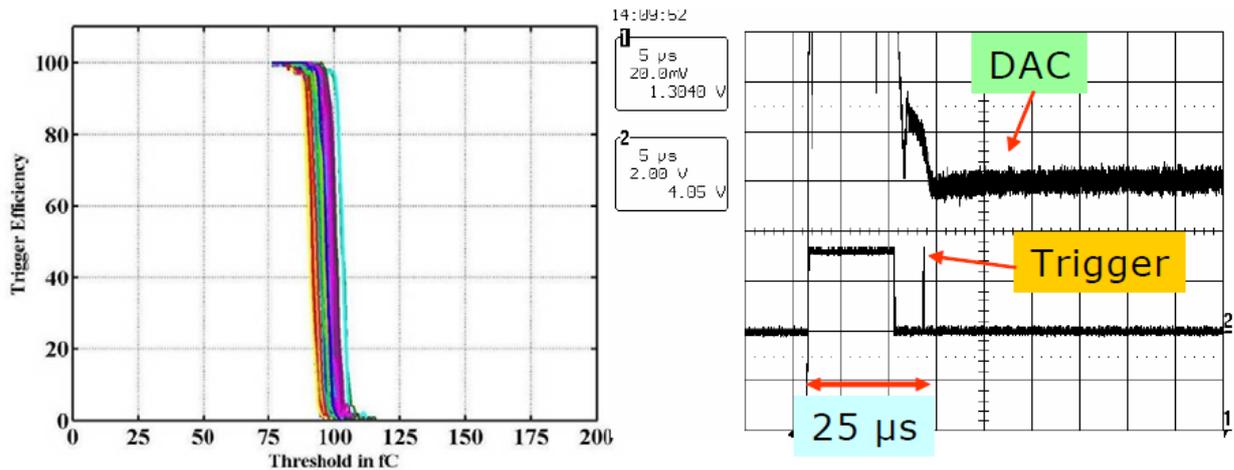

Figure 41: Trigger efficiency for 100 fC input as a function of DAC threshold.

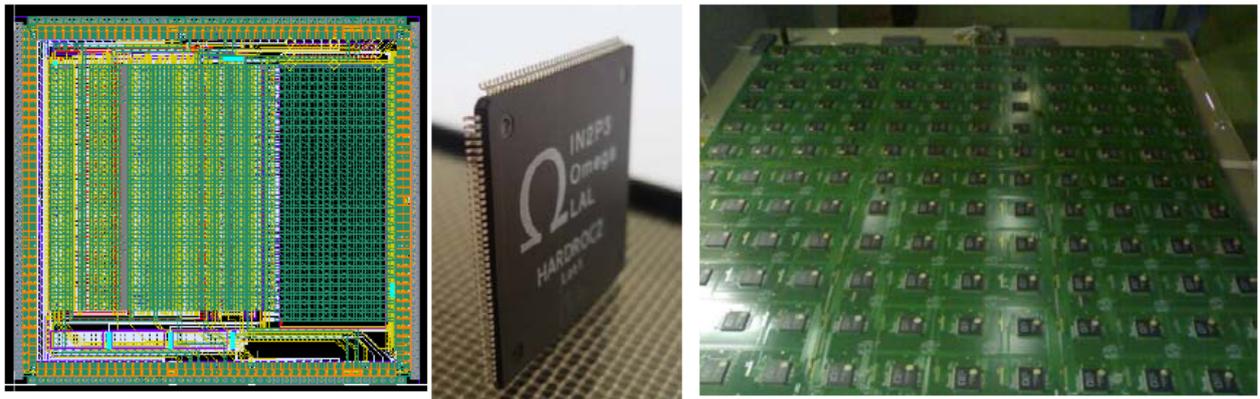

Figure 42: Layout of HARDROC2, view of the chip packaged in TQFP160 and square meter prototype of RPC DHCAL with 144 HaRDROC

The above mentioned production of 10 000 chips was launched to equip 40 Glass RPC (GRPC) planes of a one cubic meter detector that will be tested soon.

A specific test bench with a Robot (Figure 43) has been used at Lyon to test and qualify the chips before mounting them on the boards. The yield is about 93%.

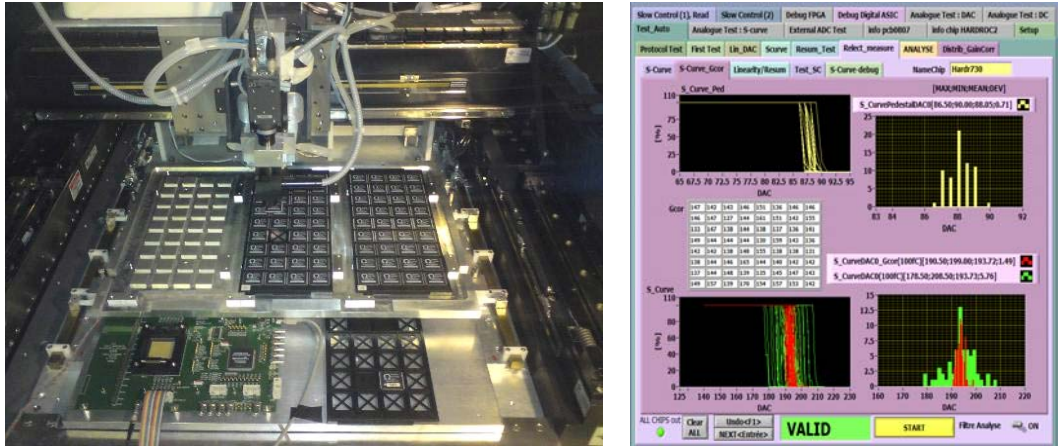

Figure 43: Testbench used to test the 10 000 HARDROC2 chips and test measurement results

**Silicon Photomultiplier Read Out Chip (SPIROC) for the Analogue Hadronic Calorimeter (AHCAL)**

The SPIROC ASIC that reads 36 SiPMs is an evolution of the FLC_SiPM used in the physics prototype. The first prototype was produced in AMS SiGe 0.35 μm. and packaged in a CQFP240 package. Similarly to HARDROC, a second version, SPIROC2, was realized to accommodate a thinner TQFP208 package and fix a bug in the ADC.

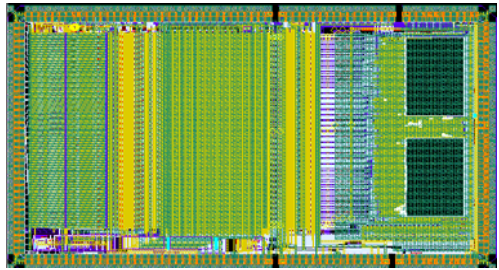

Figure 44: Layout of SPIROC2

Each channel of SPIROC2 (Figure 44 and Figure 45) is made of:

- An 8-bit input DAC with a very low power of 1 μW/channel as it is not power pulsed in order to keep the SiPM bias constant. The DAC also has the particularity of being powered with 5 V whereas the rest of the chip is powered with 3.5 V.
- A high gain and a low gain preamp in parallel on each input allow handling the large dynamic range. A gain adjustment over 4 bits common for the 36 channels has been integrated in SPIROC2. A variant (SPIROC2B) with individual gain adjustment over 6 bits has also been produced.
- The charge is measured on both gains by a "slow" shaper (50-150 ns) followed by an analog memory with a depth of 16 capacitors.
- The auto-trigger is taken on the high gain path with a high-gain fast shaper followed by a low offset discriminator. The discriminator output is used to generate the hold on the 36 channels. The threshold is common to the 36 channels, given by a 10-bit DAC similar to the one from HARDROC with a subsequent 4-bit fine tuning per channel.
- The discriminator output is also used to store the value of a 300 ns ramp in a dedicated analog memory to provide time information with an accuracy of 1ns
- A 12-bit Wilkinson ADC is used to digitize the data at the end of the acquisition period.

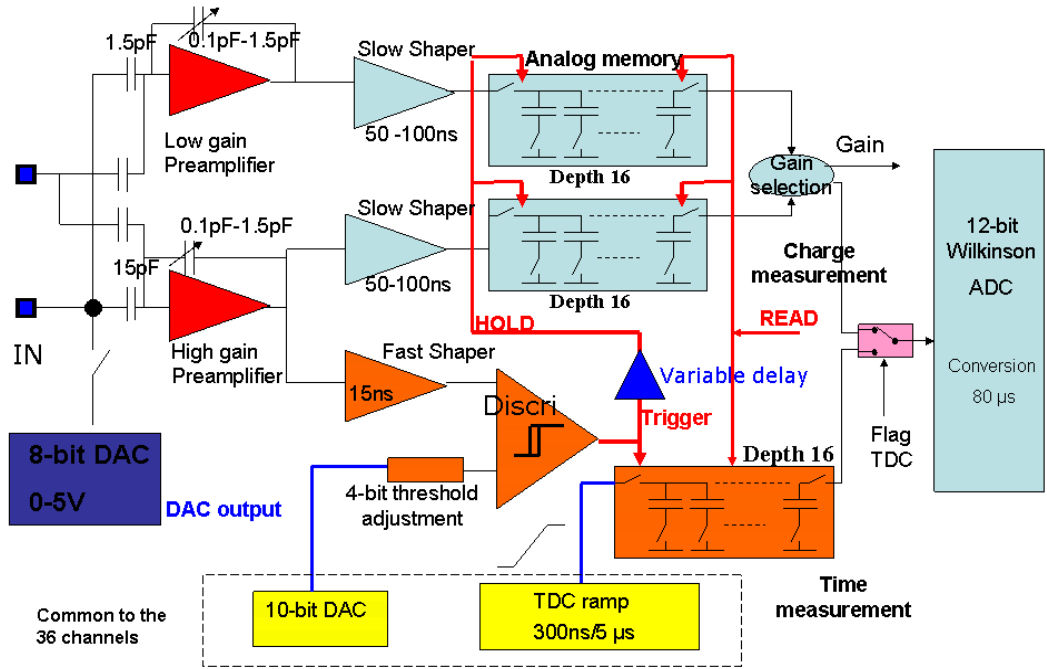

Figure 45: Schematic diagram of one channel of SPIROC2

The digital part is complex as it must handle the SCA write and read pointers, the ADC conversion, the data storage in a RAM and the readout process.

The chip has been extensively tested by many groups. The first series of tests has been mostly devoted to characterizing the analog performance, which meets the design specifications. A single photoelectron spectrum using the full chain and a LED pulser is displayed in Figure 31.
The one photo-electron signal to noise ratio is around 8.
The linearity as a function of the input charge in the auto gain mode and using the internal ADC is shown in Figure 46.

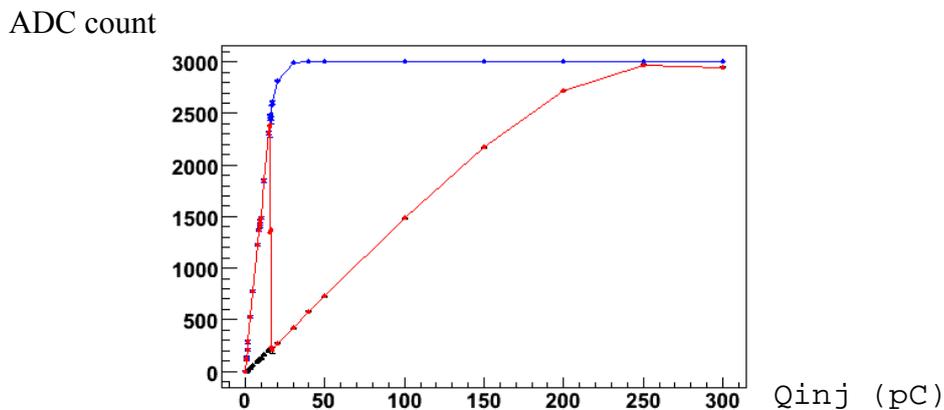

Figure 46: SPIROC2B linearity using the auto gain mode.

The digitization part has also been characterized and the 12 bit ADC exhibits a very good integral non-linearity of 1 LSB and a noise comprised between 0.5 and 1 LSB (Figure 47).

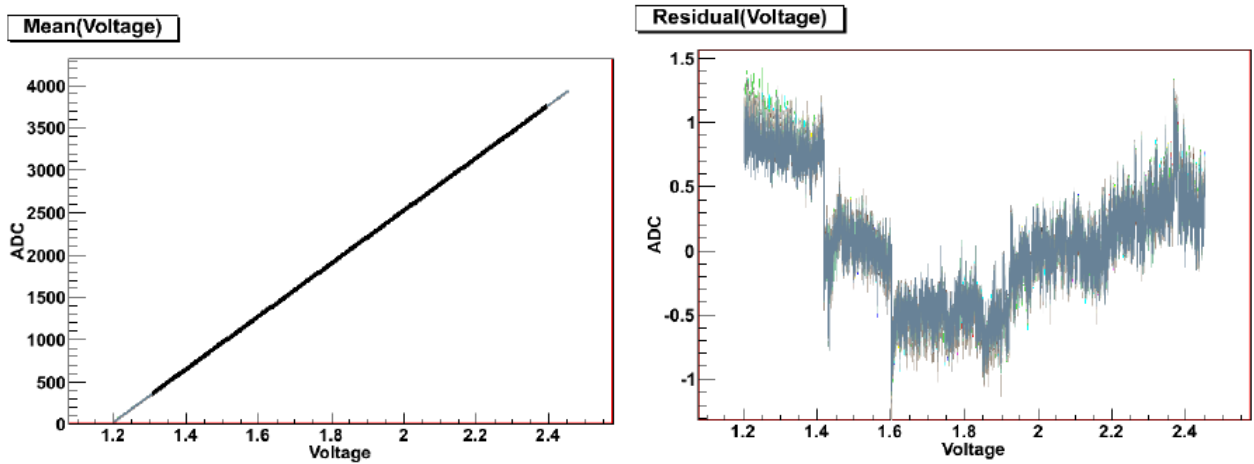

Figure 47: ADC response of the 36 channels over the 12 bits dynamic range (left). Residual to a linear fit showing an integral linearity better than 1 LSB (right).

The chips have been assembled on a HCAL PCB (HBU) and tested with detector, as described in section 5.2.3 and are being operated with good results.

**ECAL Read-Out chip**

For the ECAL, the chip SKIROC2 has been designed (Figure 48) and submitted. It keeps most of the analog part of SPIROC2, except for the preamp which is a low noise charge preamp followed by a low gain and a high gain slow shaper to handle a large dynamic range from 0.5 MIP (2 fC) up to 2500 MIPs (10 pC).

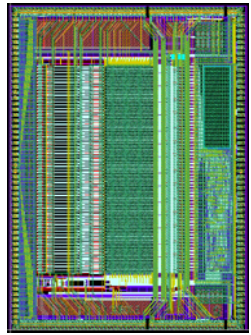

Figure 48: layout of SKIROC2 (7.2mmx8.6 mm)

SKIROC2 chips are not packaged (except for test bench measurements) as they must be directly bonded on the printed circuits. The characterization is performed on testbench using a few packaged chips but the produced chips will be tested using a probe station. FEV boards hosting 16 SKIROC2 chips are under design to readout 1024 channels.

## 1.20 Data Acquisition (DAQ)

The second version of the EUDET DAQ [56], aims at reading the direct numerical output from the technological prototypes of the CALICE collaboration at beam tests. The project is now in its penultimate phase with the integration of all elements. An up-to-date information can be found in [57-58]. It has recently triggered interest of the FCAL collaboration for a common interface, and should be part of the «common DAQ» package of the FP7 funded initiative AIDA, together with the DAQ of the telescope EUDAQ.

**Functional Specifications**

All the technological prototypes feature embedded electronics in the form of ASICs with built-in memory situated inside the detectors. This electronics performs an analogical amplification and shaping, the digitization and an internal triggering, a local storage of the data in memory.

The DAQ should handle:

- The loading of the configuration in the ASICs (seen as one stream of bits for one ASICs partition), and eventually the verification of the loaded data.
- The management of the acquisition modes through "block transfer" (ASIC and cards configuration loading) and "fast commands" concerning the acquisition running mode (single event, ILC mode), startreadout, stopreadout, reset and synchronization.
- The distribution of the fast signals in a synchronous manner: Clock, Trigger from the Clock and Control Card (CCC) to the DIFs and the collection and OR'ing of the busy signals the other way around. "Synchronous" here is detector and reconstruction specific: for the Semi-Digital HCAL (SDHCAL) a machine clock period (200 ns) should suffice, for the AHCAL the requirement is typically less than 1 ns.
- The data flux, which is here reasonably low. For the SDHCAL prototype a 100 GeV pion on the average creates hits in 4.8 HARDROC ASICs per plane. With a readout clock of 2.5 MHz, the complete data flux expected in beam tests would be of the order of 20 MB/s allowing for a maximum allowed event frequency of 3.2 kHz (much higher than the working region of standard GRPC, limited to a few 100 Hz for example). For the ECAL, equipped with 2 ASIC readout lines per DIF, 4 touched ASICs (worst case) in each plane would generate a data flux of 113 MB/s for a maximum acquisition rate of 1.2 kHz. Under the same hypothesis as for the SDHCAL, the AHCAL data flux reaches 338 MB/s, allowing for a 5 kHz event rate. In all cases, the data flux is limited by the readout speed of the ASICs.
- Two running modes: a single event mode, in which an external signal (typically signalling the passage of a particle of a given type, with a combination of scintillators and Cherenkov detectors) triggers a suspension of the acquisition mode of the ASIC, their readout and the resumption of the acquisition. An ILC-like mode, in which the ASICs are switched in Acquisition mode at the start of a particle spill, and are readout either at the end of the spill, after a given time span or number of particles, or when any of the detector ASICs emits a RAM-full signal. The acquisition is eventually resumed as soon as the readout is completed. Both modes using the auto-triggering capacity of the ASICs, they are very sensitive to the noise. A careful online monitoring of the noise has then to be performed, with eventually an automatic correction procedure to kill noisy channels.
- Connection and provide simple logic and a 50 MHz clock.

**Hardware Availability**

All the generic hardware elements, the LDA and its mezzanine cards, Data Concentration Card (DCC), Clock and Control Card (CCC), Off-Dertector Receiver (ODR) and PC, are ready and tested. A list of equipment is available in [59]. Most parts of the equipment have been dispatched to main testing labs.

The last physical parts produced were the DCC and LDA cards; 20 DCC cards have been produced and been tested, with almost no failure found, they will be installed in a standard VME crate. The LDA are physically composed of a commercial baseboard, and 3 custom mezzanine cards respectively supporting the 10 HDMI connectors, the connection to the CCC and the Gigabit-Ethernet connection. All the Hard Ware (HW) parts are available; a mechanical structure is under study to host securely the CCC and the LDAs, both having non standard formats, in the test beam area.

The last missing elements are the HDMI cables, whose features are conditioned for their length by the physical implementation of the detectors in the test beam hall and for their composition by security (halogen free cables are requested for work in CERN). Some studies just started on the mechanical implementation of the cabling for the SDHCAL; 120–150 cables of 4 to 5m are needed, another 30 for the ECAL, and 40 for the AHCAL. A survey has started to find some at a reasonable cost.

**Firmware & Implemented Functionalities**

The major occupation of the last year has concerned the Firm Ware (FW) of all the cards.

First a common FW for all DIF is being developed; a global framework was developed for the interconnection of existing various parts: an 8-bit/10-bit decoding blocks, testing modules (pseudo random pattern generation for high volume of data, and echo of configuration pattern), and working implementation of the ROC chips management used for the SDHCAL over USB.

The DCC FW, including the 8-bit/10-bit parts and a multiplexing engine is complete and works as expected; the main objective here is transparency.

The LDA FW has specifications very similar to the DCC, with the "Downstream block" being replaced by a Gigabit-Ethernet one. It is now ~98% complete with only comfort functionality missing, such as a soft reset of the board, and a small instability of the incoming flux from the PC at full speed, which can easily be avoided by a small delay in the configuration sending.

The CCC FW handling trigger and busy conditions should remain flexible to adapt for various set-ups. An implementation with all the basic functions exists, the handling of the BUSY signals from the LDAs and their treatment is being improved. The CCC CPLD code was heavily modified to allow for a direct connection to the DIF implementing clock, trigger and hard coded fast command distribution to 3 DIFs [60].

**Software**

The XDAQ framework [61] for the readout of the first SDHCAL m² prototypes, using a direct USB connection to the DIF, was developed to include most components necessary for the test beam such as event building, data quality histogram building and online reconstruction via an interface to Marlin and root histograms, a Graphical User Interface (GUI) tool to manage the acquisition and display histograms, the storage of RAW data in LCIO format. A first prototype of a configuration Data Base (DB) based on the MySQL language was tested.

The basic chain PC+LDA+DCC+DIF has been tested using ad hoc Soft Ware (SW) elements based on python scripts and GUI and C libraries. The C library was interfaced with the XDAQ environment as a driver and worked immediately, allowing for data flow intensity testing. The interface to the CCC is already done.

## Summary

The EUDET project has played an important role to maintain and extend Europe's position in advanced detector R&D required for the linear collider and beyond.

An important aim of the project was to create a network of European institutes for exchange of information as well as for performance of simulations and analysis. The provision of additional computing hardware and common software helped to realize this. Dedicated storage and computing resources for EUDET was set up at the Bonn University, DESY and Tel Aviv University.

A high precision ($\sigma < 2$ μm) and fast pixel telescope has been built for tests of various detector elements at the DESY and CERN test beams. A flexible and user-friendly DAQ system has been developed to satisfy the various needs of the users. The telescope was frequently used by different groups testing different types of detectors also outside the EUDET-collaboration.

An infrastructure for testing various technologies used in gaseous tracking devises as well as for Si-tracking has been built up. This comprises the construction of a large TPC field cage together with an endplate allowing seven gas amplification modules to be mounted. It is supplemented by a general purpose readout electronics and DAQ. The field cage is placed in a large bore superconducting magnet, which provides a maximum field of 1.2T. This infrastructure has been used to test GEM and Micromegas structures together with pad and Si-pixel readout. Test results have shown that the required resolution in space can be achieved. Several tests with the Silicon test infrastructure together with the EUDET telescope have been performed and valuable results have been obtained.

For the calorimeter development, the EUDET initiative was instrumental in performing the step towards realistic second generation technological prototypes, defining the read-out architecture and a homogenous ASIC development for different electromagnetic and hadronic sensor options. These have been successfully validated with various demonstrator prototypes. The resulting structure form an excellent starting point for the follow-up program AIDA, in which the prototypes will be extended to full size, and their use be expanded into studies for applications in the multi-TeV region, where the shower timing behaviour becomes important.

## Acknowledgement

This work is supported by the Commission of the European Communities under the 6th Framework Programme "Structuring the European Research Area", contract number RII3-026126.

## References

1. F.Gaede, J.Engels, "Marlin et al - A Software Framework for ILC detector R&D", EUDET-Report-2007-11
2. http://ilcsoft.desy.de/portal/software_packages/gear#
3. F.Gaede, T.Behnke, N.Graf, T.Johnson, "CHEP03 March24-28, 2003 La Jolla, USA", TUKT001, arXiv:physics/0306114
4. F.Gaede, ``Marlin and LCCD: Software tools for the ILC,'' Nucl.Instrum.Meth. A 559, (2006) 177

5. http://ilcsoft.desy.de
6. EUDET-MEMO-2010-15
7. R. Turchetta et al., "A monolithic active pixel sensor for charged particle tracking and imaging using standard VLSI CMOS technology", Nucl. Instr. & Meth. in Phys. Res. Sect. A 458 (2001) 677-689.
8. M. Winter, "CMOS Pixel Sensors for Charged Particle Tracking: Achieved Performances and Perspectives", 1st international conference on Technology and Instrumentation in Particle Physics, 2009, Tsukuba, Japan.
9. A. Cotta Ramusino, EUDET-Memo-2007-36
10. http://www.lctpc.org
11. The International Large Detector, Letter of Intent, The ILD Concept Group, DESY-2009/87, FERMILAB-PUB-09-682-E, KEK Report 2009-6
12. C.Grefe, ``Magnetic field map for a large TPC prototype'', Hamburg. Univ., FB Physik, Dipl.arb., 2008
13. http://ilcsoft.desy.de/portal/software_packages/marlintpc/
14. T. Behnke et al., A lightweight fieldcage for a large TPC prototype for the ILC, JINST 5 (2010) P10011, arXiv:1006.3220
15. D. Peterson, http://www.lepp.cornell.edu/~dpp/linear\_collider/LargePrototype.html
16. Y. Giomataris et al., "Micromegas: A High Granularity Position Sensitive Gaseous Detector for High Particle Flux Environments", Nucl. Instrum. Meth. A376(1996)29
17. M. Dixit et al., Nucl. Instrum. Meth. A518(2004)721
18. F. Sauli, "GEM: A New Concept for Electron Amplification in Gas Detectors", Nucl. Instrum. Meth. A386(1997)531
19. L.Joensson, U.Mjoernmark, 'Front-end electronics and data acquisition for the LCTPC', Eudet-Memo-2007-53
20. A.Oskarsson et al., 'A General Purpose Electronic readout system for tests of Time Projection Chambers, equipped with different avalanche multiplication systems', Eudet-Memo-2008-49
21. L.Musa, 'Prototype compact readout system', Eudet-Memo-2009-31
22. L.Joensson on behalf of the LCTPC collaboration, 'Front end electronics for a TPC at future linear colliders', Eudet-Memo-2010-30
23. A. Kaukher, Ph.D. thesis, University of Rostock, 2008
24. P. Baron et al., "AFTER, an ASIC for the Readout of the Large T2K Time Projection Chambers", IEEE TNS vol. 55, Issue 3, Part 3, June 2008, pp. 1744-1752
25. A.Ishikawa et al., "A GEM TPC End-Panel Pre-Prototype", arxiv.org/abs/0710.0205
26. M. Ljunggren on behalf of the LCTPC collaboration, 'Analysis of data recorded with the LCTPC equipped with a two layer GEM system', Proc. of the TIPP 2011 conference, Chicago, USA, June 9-14, 2011.
27. V.Blobel and C.Kleinwort, "A New Method for the High-Precision Alignment of Track Detectors", arxiv.org/abs/hep-ex/0208021
28. D. Karlen, "Measuring Distortions in a TPC with Photoelectrons". In the Proceedings of 2007 International Linear Collider Workshop (LCWS07 and ILC07), Hamburg Germany
29. P. Colas et al., "The readout of a GEM- or micromegas-equipped TPC by means of the Medipix2 CMOS sensor as direct anode", Nucl. Instrum. Meth. A535: 506-510, 2004
30. M. Campbell et al., "The Detection of single electrons by means of a micromegas-covered MediPix2 pixel CMOS readout circuit", Nucl. Instrum. Meth. A540: 295-304, 2005
31. M. Chefdeville et al., "An electron-multiplying 'Micromegas' grid made in silicon wafer post-processing technology", Nucl. Instrum. Meth. A556: 490-494, 2006
32. V.M. Blanco Carballo et al., "A radiation imaging detector made by postprocessing a

standard CMOS chip", IEEE Device Letters 29: 585-587, 2008
33. X. Llopart et al., "Medipix2: A 64-k pixel readout chip with 55-μm square elements working in single photon counting mode", IEEE Trans. Nucl. Sci. NS-49: 2279-2283, 2002
34. X. Llopart et al., "Timepix, a 65k programmable pixel readout chip for arrival time, energy and/or photon counting measurements", Nucl. Instrum. Meth. A581: 485-494, 2007; Erratum ibid. A585: 106-108, 2008
35. A. Bamberger et al., "Readout of GEM Detectors Using the Medipix2 CMOS Pixel Chip", Nucl. Instrum. Meth. A573: 361-370, 2007
36. A. Bamberger et al., "Resolution studies on 5-GeV electron tracks observed with triple-GEM and MediPix2/TimePix-readout", Nucl. Instrum. Meth. A581: 274-278, 2007
37. Y. Bilevych et al., "Spark protection layers for CMOS pixel anode chips in MPGDs", Nucl. Instrum. Meth. A629: 66-73, 2011
38. Y. Bilevych et al., "TwinGrid: A wafer post-processed multistage micro patterned gaseous detector", Nucl. Instrum. Meth. A610: 644-648, 2009
39. V.M. Blanco Carballo et al., "GEMGrid: a wafer post-processed GEM-like radiation detector", Nucl. Instrum. Meth. A608: 86-91, 2009
40. T.H. Pham et al., A 130nm CMOS mixed mode front end readout chip for silicon strip tracking at the future linear collider, in Nucl.Instrum.Meth.A623:498-500,2010
41. www.umc.com/English
42. www.gm-ideas.com
43. C. Foudas et al., "The CMS Tracker Readout Front End Driver", arXiv: physics/0510229
44. M. Fernandez et al., 'Semitransparent microstrip detectors for infrared laser alignment of particle trackers', EUDET-Memo-2010-20
45. D. Bassignana et al., Silicon microstrip detectors for future tracker alignment systems, in Nucl.Instrum.Meth.A628:276-281,2011
46. M. French et al., 'Design and results from the APV25, a deep sub-micron CMOS front-end chip for the CMS tracker', Nucl. Instr. and Meth. A
47. M. Friedl: 'The CMS silicon strip tracker and its electronic readout', PhD thesis (2004);
48. EUDET-Report-2009-01
49. L. Raux et al., "SPIROC Measurement: Silicon Photomultiplier Integrated Readout Chips for ILC", Proc. IEEE Nuclear Science Symposium (NSS08), Dresden, Germany, 2008
50. K. Gadow et al., "Concept, Realization and Results of the Mechanical and Electronics Integration Efforts for an Analog Hadronic Calorimeter", EUDET-Report-2010-02.
51. M. Wing et al., "Calorimeter DAQ status", EUDET-Memo-2009-27
52. I. Polak et al.,"An LED calibration system for the CALICE HCAL", Proc. IEEE Nuclear Science Symposium (NSS10), Knoxville, Tennessee, USA, Nov. 2010
53. J.Cvach et al., "Magnetic field tests of the QRLED driver", EUDET-memo-2009-05
54. M. Idzik, Sz. Kulis, D. Przyborowski, "Development of front-end electronics for the luminosity detector at ILC", Nucl. Instr. and Meth. A, vol. 608, 2009
55. M. Idzik, K. Swientek, T. Fiutowski, S. Kulis and P. Ambalathankandy, 'A power scalable 10-bit pipeline ADC for Luminosity Detector at ILC', JINST 6 (2011) P01004
56. M.J. Goodrick et al., 'Development of a modular and scalable data acquisition system for calorimeters at a linear collider', JINST 6 (2011) P10011.
57. https://twiki.cern.ch/twiki/bin/view/CALICE/CALICEDAQ
58. http://ilcagenda.linearcollider.org/categoryDisplay.py?categId=154
59. https://twiki.cern.ch/twiki/bin/view/CALICE/HardwareList
60. J. Prast talk: http://ilcagenda.linearcollider.org/conferenceDisplay.py?confId=4391

61. https://svnweb.cern.ch/trac/cmsos